\documentclass[12pt,letterpaper,english,draftclsnofoot, onecolumn]{IEEEtran}
\usepackage[T1]{fontenc}
\usepackage[latin9]{inputenc}
\usepackage{amsmath}
\usepackage{amssymb}
\usepackage{stmaryrd}
\usepackage{graphicx}
\usepackage{esint}

\makeatletter

\providecommand{\tabularnewline}{\\}

\usepackage{cite}
\usepackage{algorithm}
\usepackage{algorithmicx}
\usepackage{algpseudocode}
\usepackage{babel}

\allowdisplaybreaks[4]
\usepackage{flushend}

\@ifundefined{showcaptionsetup}{}{%
 \PassOptionsToPackage{caption=false}{subfig}}
\usepackage{subfig}
\makeatother

\usepackage{babel}
\begin{document}

\title{Message-Passing Receiver for Joint Channel Estimation and Decoding
in 3D Massive MIMO-OFDM Systems}

\author{Sheng Wu, \IEEEmembership{Member, IEEE}, Linling Kuang, \IEEEmembership{Member, IEEE},\\
 Zuyao Ni, Defeng (David) Huang, \IEEEmembership{Senior Member, IEEE},\\
Qinghua Guo, \IEEEmembership{Member, IEEE}, and Jianhua Lu, \IEEEmembership{Fellow, IEEE}\thanks{This work was partially supported by the National Nature Science Foundation
of China (Grant Nos. 91338101, 61231011, and 91438206), the National
Basic Research Program of China (Grant No. 2013CB329001), and Tsinghua
University Initiative Scientific Research Program (Grant No. 20131089219).

Sheng Wu, Linling Kuang and Zuyao Ni are with the Tsinghua Space Center,
Tsinghua University, China (e-mail: %
\mbox{%
\{thuraya, kll, nzy\}%
}@tsinghua.edu.cn).

Defeng (David) Huang is with the School of Electrical, Electronic
and Computer Engineering, The University of Western Australia, Australia
(e-mail: %
\mbox{%
david.huang%
}@ee.uwa.edu.au).

Q. Guo is with the School of Electrical, Computer and Telecommunications
Engineering, University of Wollongong, Australia, and is also with
the School of Electrical, Electronic and Computer Engineering, The
University of Western Australia, Australia (e-mail: %
\mbox{%
qinghua.guo%
}@uow.edu.au).

Jianhua Lu is with the Department of Electronic Engineering, Tsinghua
University, China (e-mail: %
\mbox{%
lhh-dee%
}@mail.tsinghua.edu.cn).}}
\maketitle
\begin{abstract}
In this paper, we address the message-passing receiver design for
the 3D massive MIMO-OFDM systems. With the aid of the central limit
argument and Taylor-series approximation, a computationally efficient
receiver that performs joint channel estimation and decoding is devised
by the framework of expectation propagation. Specially, the local
belief defined at the channel transition function is expanded up to
the second order with Wirtinger calculus, to transform the messages
sent by the channel transition function to a tractable form. As a
result, the channel impulse response (CIR) between each pair of antennas
is estimated by Gaussian message passing. In addition, a variational
expectation-maximization (EM)-based method is derived to learn the
channel power-delay-profile (PDP). The proposed joint algorithm is
assessed in 3D massive MIMO systems with spatially correlated channels,
and the empirical results corroborate its superiority in terms of
performance and complexity.\end{abstract}


\section{Introduction}

Recently, massive multiple-input multiple-output (MIMO) systems with
tens to hundreds of antennas at the base-station (BS) have gained
significant attention \cite{Marzetta10,Rusek2011ScalingUpMIMO,Hoydis2013,Larsson2014}.
It has been proved that massive MIMO systems can scale down transmit
power as well as increase spectrum efficiency by orders of magnitude
\cite{Rusek2011ScalingUpMIMO}. One of the tasks in massive MIMO systems
is estimating the channel impulse response (CIR) for each transmit-receive
link, since high data rates and energy efficiency can only be achieved
when CIR is known \cite{Shariati2014}. In contrast to the conventional
MIMO systems employing a small number of antennas, there are a large
number of channels need to be estimated. The pilot overhead required
for channel estimation is proportional to the number of transmit antennas,
which can be excessive in massive MIMO systems \cite{Noh2014Pilot}.
In the meantime, the available resources for training are restricted
by the channel coherence time. On the other hand, the energy consumption
by baseband processing grows with the number of antennas, which may
obliterate the advantage of massive MIMO systems in energy efficiency.
Thus, low-complexity channel estimation with high accuracy and reduced
overhead is critical to massive MIMO systems.

Iterative receivers that jointly estimate the channel coefficients
and detect the data symbols are able to provide more accurate channel
estimation with less training overhead \cite{salvo2008joint,novak2013idma,LiuYinsheng2014,ZhangPeichang2014,Ma2014,Park2015}.
Factor graph and sum-product algorithm (SPA) \cite{Kschischang2001}
have been used as a unified framework for iterative joint data detection,
channel estimation, interference cancellation, and decoding \cite{worthen2001unified,Wymeersch2007}.
However, exact SPA for joint channel estimation and decoding is computationally
infeasible. To overcome this problem, various message-passing algorithms
based on approximate inference have been proposed \cite{liu2009joint,Kirkelund2010,Carles2011,guo2011based,schniter2011message,knievel2012multi,novak2013idma,Riegler2013,ZhangXiaoying2014}.
In existing approaches, the message passing strategies include loopy
belief propagation (LBP) \cite{liu2009joint,guo2011based,schniter2011message,knievel2012multi,novak2013idma},
variational methods \cite{lin2009variational,Kirkelund2010,ZhangXiaoying2014},
and a hybrid of both \cite{Carles2011,Riegler2013}.

LBP has a high complexity when applied to graphical models that involve
both discrete and continuous random variables. This has been addressed
by merging the SPA with the expectation-maximization (EM) algorithm
\cite{guo2011based} or approximating the messages of SPA with Gaussian
messages \cite{schniter2010joint,schniter2011message,guo2011based,novak2013idma}.
Variational inference methods have been applied to MIMO receivers
for joint detection, channel estimation, and decoding \cite{Kirkelund2010}.
In \cite{Carles2011}, Riegler \emph{et al.} derived a generic message-passing
algorithm that merges belief propagation (BP) with the mean-field
(MF) approximation (BP-MF), and applied it to joint channel estimation
and decoding in single-input single-output orthogonal frequency division
multiplex (OFDM) systems and MIMO-OFDM systems \cite{Carles2011,Badiu2012,Riegler2013}.
The BP-MF has to learn the noise precision to take into account the
residual interference from other users even when the noise power is
known \cite{Dremeau2012,Krzakala2014}, as the channel transition
functions are incorporated into the MF part \cite{Carles2011,Badiu2012,Riegler2013}.
Otherwise, the uncertainty of residual interference is completely
ignored, and the likelihood function associated with the messages
extracted from observations tends to overwhelm the\emph{ a priori
}probability. Besides, the BP-MF requires high computational complexity
as large matrices need to be inverted to estimate channel frequency
response (CFR) \cite{Carles2011}, and thereby it is only feasible
in the case of a few antennas and subcarriers. We note that there
is a low-complexity version of the BP-MF algorithm proposed in \cite{Badiu2013},
but its performance is inferior. The degraded performance may be due
to the unrealistic assumption that groups of contiguous channel weights
in frequency-domain obey a Markov model.

To achieve joint channel estimation and decoding for massive MIMO
systems using OFDM modulation in frequency-selective channels, the
receiver needs to complete three tasks: decoupling frequency-domain
channel coefficients and data symbols from noisy observations, decoding,
and channel estimation. Via central-limit theorem and moment matching,
an approximate BP has been derived in \cite{novak2013idma}, \cite{liu2009joint}
and \cite{knievel2012multi}. Despite its superior performance, the
approximate BP bears a heavy computation burden: it needs to take
a large number of moment-matching operations, each being highly complicated.
In this paper, we use the framework of expectation propagation (EP)
\cite{minka2001family} to derive an efficient message-passing algorithm.
Specifically, at the channel transition functions, we use the central-limit
theorem to efficiently obtain the beliefs of frequency-domain channel
coefficients and the beliefs of data symbols, and then employ a quadratic
approximation to project them into the Gaussian family. In the meantime,
the expectation propagation principle is applied to the symbol-variable
nodes. As the beliefs of frequency-domain channel coefficients are
now in the form of Gaussian family, a Gaussian message passing based
estimator \cite{Wu1405:Expectation} can be employed, which exploits
the fact that the CFR is the Fourier transformation of the CIR. Furthermore,
using the beliefs of time-domain channel taps, the unknown power-delay-profile
(PDP) can be learned by variational expectation maximization. We note
that Parker \emph{et al}. applied central-limit theorem and Taylor-series
approximations to formulate a bilinear generalized approximate message-passing
algorithm for the SPA in the high dimensional limit \cite{Parker2014Bilinear},
but its scope is different from that of this work.

The proposed scheme of joint channel estimation and decoding is assessed
in 3D massive MIMO systems with spatially correlated channels. Experiments
show that its performance is within 1 dB of the known-channel bound
in both a $64\times8$ MIMO system and a $16\times8$ MIMO system,
and outperforms the performance of BP-MF by 0.4 dB in the $16\times8$
MIMO system, the low-complexity version of BP-MF by 1.2 dB in the
$64\times8$ MIMO system and 1.6 dB in the $16\times8$ MIMO system.
On the other hand, the complexity of the proposed algorithm is a small
percentage of that of BP-MF and $\frac{1}{3}$ of that of the low-complexity
version of BP-MF.

The remainder of this paper is organized as follows. The system model
is described in Section \ref{sec:System-Model}. In Section \ref{sec:Hybrid-Message-Passing}
the message passing for joint detection and decoding is detailed.
Complexity comparisons are shown in Section \ref{sec:Complexity-Comparisons},
and numerical results are provided in Section \ref{sec:Simulation-Results},
followed by conclusions in Section \ref{sec:Conclusion}.

\emph{Notation}: Lowercase letters (e.g., $x$) denote scalars, bold
lowercase letters (e.g.,$\boldsymbol{x}$) denote column vectors,
and bold uppercase letters (e.g., $\boldsymbol{X}$) denote matrices.
The superscripts $(\cdot)^{\mathsf{T}}$, $(\cdot)^{\mathsf{H}}$
and $(\cdot)^{*}$ denote the transpose operation, Hermitian transpose
operation, and complex conjugate operation, respectively. Also, $\mathsf{diag}\{\boldsymbol{x}\}$
denotes a square diagonal matrix with the elements of vector $\boldsymbol{x}$
on the main diagonal; $\boldsymbol{X}\otimes\boldsymbol{Y}$ denotes
Kronecker product of $\boldsymbol{X}$ and $\boldsymbol{Y}$; $\ensuremath{\boldsymbol{I}}$
denotes an identity matrix; and $\mathsf{ln}(\cdot)$ denotes the
natural logarithm. Furthermore, $\mathcal{N}_{\mathbb{C}}(x;\hat{x},\nu_{x})=(\pi\nu_{x})^{-1}\exp(-\left|x-\hat{x}\right|^{2}/\nu_{x})$
denotes the Gaussian probability density function (PDF) of $x$ with
mean $\hat{x}$ and variance $\nu_{x}$; and $\mathsf{Gam}(\gamma;\alpha,\beta)=\beta^{\alpha}\gamma^{\alpha-1}\exp(-\beta\gamma)/\Gamma(\alpha)$
denotes the Gamma PDF of $\gamma$ with shape parameter $\alpha$
and rate parameter $\beta$, where $\Gamma(\cdot)$ is the gamma function.
Finally, $\propto$ denotes equality up to a constant scale factor;
$\boldsymbol{x}\backslash x_{tnk}$ denotes all elements in $\boldsymbol{x}$
but $x_{tnk}$; and $\mathsf{E}_{p(x)}\{\cdot\}$ denotes expectation
with respect to distribution $p(x)$.

\section{System Model \label{sec:System-Model}}

We consider the uplink of a massive MIMO system where $N$ single
antenna users communicate with a BS simultaneously. The BS employs
a uniform planar array (UPA) consisting of $M=(D\times W)\gg N$ antennas
distributed across $D$ rows and $W$ columns. Frequency-selective
block-fading channels are assumed, and OFDM is employed to combat
multipath interference.

\subsection{Channel Model}

The CIR between the $n\text{th}$ user and the $m\text{th}$ receive
antenna is denoted by $\boldsymbol{h}_{mn\cdot}=[h_{mn1}\cdots h_{mnL}]^{\mathsf{T}}$,
where $h_{mnl}$ is the $l\text{th}$ path gain and $L$ is the maximum
channel spread. Let $\boldsymbol{h}_{\cdot nl}=[h_{1nl}\cdots h_{Mnl}]^{\mathsf{T}}$
denote gain vector of the $l\text{th}$ paths between user $n$ and
all the $M$ receive antennas at the BS. Due to close antenna spacing
at the BS, we can assume that the $M$ CIRs between the user $n$
and all the $M$ receive antennas at the BS follow an identical PDP
$\{\mathsf{E}\{|h_{mnl}|^{2}\}\triangleq\alpha_{nl},\forall m\}$.
We can also assume that the transmit antennas from different users
are spatially uncorrelated. Accordingly, the Kronecker spatial fading
correlation model for the gain vector $\boldsymbol{h}_{\cdot nl}$
is given by \cite{Schumacher2002MIMO}
\begin{equation}
\boldsymbol{h}_{\cdot nl}=\boldsymbol{R}_{nl}^{\frac{1}{2}}\boldsymbol{h}_{nl}^{\mathsf{iid}},\label{eq:kronecker_channel}
\end{equation}
where $\boldsymbol{R}_{nl}\in\mathbb{C}^{M\times M}$ denotes the
receive correlation matrix, and $\boldsymbol{h}_{nl}^{\mathsf{iid}}\in\mathbb{C}^{M\times1}$
denotes independent complex Gaussian matrix with zero mean and covariance
matrix $\alpha_{nl}\boldsymbol{I}$. A ray-based 3D channel model
from \cite{Ying2014} is employed, and the receive correlation matrix
$\boldsymbol{R}_{nl}$ is approximated by
\begin{equation}
\boldsymbol{R}_{nl}\approx\boldsymbol{R}_{nl}^{\mathsf{az}}\otimes\boldsymbol{R}_{nl}^{\mathsf{el}},\label{eq:kronecker_approx}
\end{equation}
where $\boldsymbol{R}_{nl}^{\mathsf{az}}\in\mathbb{R}^{W\times W}$
and $\boldsymbol{R}_{nl}^{\mathsf{el}}\in\mathbb{R}^{D\times D}$
are the correlation matrices in azimuth and elevation directions,
respectively, and are defined by \cite{Ying2014}
\begin{align}
[\boldsymbol{R}_{nl}^{\mathsf{az}}]_{ww'} & =\frac{1}{\sqrt{b}}\mathsf{exp}\left(-\frac{a^{2}\mathsf{cos}^{2}(\theta_{nl}^{\mathsf{az}})-2jc\mathsf{cos}(\theta_{nl}^{\mathsf{az}})+\nu_{nl}^{\mathsf{az}}\left(c\mathsf{sin}(\theta_{nl}^{\mathsf{az}})\right)^{2}}{2b}\right),\\{}
[\boldsymbol{R}_{nl}^{\mathsf{el}}]_{dd'} & =\mathsf{exp}\left(2\frac{j\pi\lambda d^{\mathsf{el}}\left(d'-d\right)\mathsf{cos}\left(\theta_{nl}^{\mathsf{el}}\right)-\nu_{nl}^{\mathsf{el}}\left(\pi d^{\mathsf{el}}\left(d'-d\right)\mathsf{sin}\left(\theta_{nl}^{\mathsf{el}}\right)\right)^{2}}{\lambda^{2}}\right),
\end{align}
in terms of
\begin{align}
a & =\frac{2\pi d^{\mathsf{az}}}{\lambda}\sqrt{\nu_{nl}^{\mathsf{el}}}\left(w'-w\right)\mathsf{cos}\left(\theta_{nl}^{\mathsf{el}}\right),\\
b & =\nu_{nl}^{\mathsf{az}}a^{2}\mathsf{sin}^{2}\left(\theta_{nl}^{\mathsf{az}}\right)+1,\\
c & =\frac{2\pi d^{\mathsf{az}}}{\lambda}\left(w'-w\right)\mathsf{sin}\left(\theta_{nl}^{\mathsf{el}}\right),
\end{align}
where $\lambda$ is the carrier wavelength, $\theta_{nl}^{\mathsf{az}}$
and $\theta_{nl}^{\mathsf{el}}$ are the mean of horizontal angle-of-departure
(AoD) and the mean of vertical AoD, respectively; $\nu_{nl}^{\mathsf{az}}$
and $\nu_{nl}^{\mathsf{el}}$ are the variance of horizontal AoD and
the variance of vertical AoD, respectively; $d_{\mathsf{el}}$ and
$d_{\mathsf{az}}$ are the vertical antenna spacing and the horizontal
antenna spacing, respectively.

\subsection{Signal Model}

For the $n\text{th}$ user, the information bits $\boldsymbol{b}_{n}$
are encoded and interleaved, yielding a sequence of coded bits $\boldsymbol{c}{}_{n}$.
Then each $Q$ bits in $\boldsymbol{c}_{n}$ are mapped to one modulation
symbol $\boldsymbol{x}_{n}^{d}$, which is chosen from a $2^{Q}$-ary
constellation set $\mathcal{A}$, i.e., $\left|\mathcal{A}\right|=2^{Q}$.
The data symbols $\boldsymbol{x}_{n}^{d}$ are then multiplexed with
pilot symbols $\boldsymbol{x}_{n}^{p}$, forming the transmitted symbols
sequence $\boldsymbol{x}_{n}$. Pilot and data symbols are arranged
in an OFDM frame of $T$ OFDM symbols, each consisting of $K$ subcarriers.
Specifically, the frequency-domain symbols in the $t\text{th}$ OFDM
symbols transmitted by the $n\text{th}$ user are denoted by $\boldsymbol{x}_{tn\cdot}=[x_{tn1},\ldots,x_{tnK}]^{\mathsf{T}}$,
where $x_{tnk}\in\mathcal{A}$ denotes the symbol transmitted at the
$k\text{th}$ subcarrier. In each OFDM frame, there are $K_{p}\leq K$
pilot subcarriers in one selected OFDM symbol and the pilot subcarriers
are spaced uniformly. The set of pilot-subcarriers of user $n$ is
denoted by $\mathcal{P}_{n}=\{(t,k):x_{tnk}\thinspace\text{is a pilot symbol}\},\left|\mathcal{P}_{n}\right|=K_{p}$,
and the set of data-subcarriers is denoted by $\mathcal{D}=\overline{\bigcup_{n}\mathcal{P}_{n}}$.
To maintain the orthogonality between the pilot sequences sent by
different user, pilots symbols can be frequency division multiplexing,
time division multiplexing, code division multiplexing or hybrid of
them. For simplicity, the sets of pilot-subcarriers belong to different
users are set to be mutually exclusive, i.e., $\bigcap_{n}\mathcal{P}_{n}=\emptyset$,
and only one user actually transmits a pilot symbol at a given subcarrier,
whereas the other users transmit zero-symbol at this subcarrier \cite{Dai2013Spectrally},
i.e., $x_{tn'k}=0,\forall n'\neq n$, if $(t,k)\in\mathcal{P}_{n}$.
To modulate the OFDM symbol, a $K$-point inverse discrete Fourier
transform (IDFT) is applied to the symbol sequence $\boldsymbol{x}_{tn\cdot}$
and then a cyclic prefix (CP) is added before transmission.

At the receiver, the CP is removed first and then the received signal
from each receive antenna is converted into frequency domain through
a $K$-point discrete Fourier transform (DFT). It is assumed that
the $N$ transmitters and the receiver are synchronized and the duration
of the cyclic prefix is larger than the maximum delays. And then the
received signal during the interval of the $t\text{th}$ OFDM symbol
can be written as
\begin{align}
y_{tmk} & =\sum_{n}w_{mnk}x_{tnk}+\varpi_{tmk},\label{eq:receive signal model}
\end{align}
where $y_{tmk}$ denotes the received signal at the $k\text{th}$
subcarrier on the $m\text{th}$ receive antenna, $\varpi_{tmk}$ denotes
a circularly symmetric complex noise with zero mean and the variance
of $\sigma_{\varpi}^{2}$, and $w_{mnk}$ denotes the CFR at the $k\text{th}$
subcarrier between the $n\text{th}$ user and the $m\text{th}$ receive
antenna, which is given by
\begin{equation}
w_{mnk}=\sum_{l=1}^{L}h_{mnl}\mathsf{exp}\left(-\frac{j2\pi lk}{K}\right).\label{eq:w_ji^i}
\end{equation}
The received signal for a frame of $T$ OFDM symbols can be recast
in a matrix-vector form as
\begin{align}
\boldsymbol{y} & =\sum_{n=1}^{N}\boldsymbol{W}_{n}\boldsymbol{x}_{n}+\boldsymbol{\varpi}=\boldsymbol{W}\boldsymbol{x}+\boldsymbol{\varpi},
\end{align}
where $\boldsymbol{y}=[\boldsymbol{y}_{1}^{\mathsf{T}}\cdots\boldsymbol{y}_{M}^{\mathsf{T}}]^{\mathsf{T}}$
with $\boldsymbol{y}_{m}=[y_{1m1}\cdots y_{1mK}\cdots y_{Tm1}\cdots y_{TmK}]^{\mathsf{T}}$
denoting the received signal at the $m\text{th}$ receive antenna
for $T$ OFDM symbols, $\boldsymbol{W}_{n}=[\boldsymbol{I}_{T}\otimes\mathsf{diag}\{\boldsymbol{w}_{1n\cdot}\}\cdots\boldsymbol{I}_{T}\otimes\mathsf{diag}\{\boldsymbol{w}_{Mn\cdot}\}]^{\mathsf{T}}$
with $\boldsymbol{w}_{mn\cdot}=[w_{mn1}\cdots w_{mnK}]^{\mathsf{T}}$
denoting the CFR from the $n\text{th}$ user to the $m\text{th}$
antenna, $\boldsymbol{W}=[\boldsymbol{W}_{1}\cdots\boldsymbol{W}_{N}]$,
$\boldsymbol{x}=[\boldsymbol{x}_{1}^{\mathsf{T}}\cdots\boldsymbol{x}_{N}^{\mathsf{T}}]^{\mathsf{T}}$
with $\boldsymbol{x}_{n}=[x_{1n1}\cdots x_{1nK}\cdots x_{Tn1}\cdots x_{TnK}]^{\mathsf{T}}$
denoting the symbols transmitted by the $n\text{th}$ user, and $\boldsymbol{\varpi}=[\boldsymbol{\varpi}_{1}^{\mathsf{T}}\cdots\boldsymbol{\varpi}_{M}^{\mathsf{T}}]^{\mathsf{T}}$
with $\boldsymbol{\varpi}_{m}=[\varpi_{1m1}\cdots\varpi_{1mK}\cdots\varpi_{Tm1}\cdots\varpi_{TmK}]^{\mathsf{T}}$
denoting the noise signal at the $m\text{th}$ receive antenna.

\subsection{Factor Graph Representation of the Massive MIMO-OFDM Systems}

Our goal is to infer the information bits $\{\boldsymbol{b}_{n}\}$
from the observations $\boldsymbol{y}$ with the known pilot symbols
$\{\boldsymbol{x}_{n}^{p}\}$. In particular, we aim to achieve the
minimum bit error rate (BER) utilizing the maximum\emph{ a posteriori
}marginal criterion, i.e.,
\begin{equation}
\begin{aligned}\hat{b}_{n\iota}=\underset{b_{n\iota}\in\left\{ 0,1\right\} }{\mathsf{arg\thinspace max}}\thinspace p\left(b_{n\iota}\mid\boldsymbol{y}\right),\end{aligned}
\end{equation}
where $b_{n\iota}$ denotes the $\iota\text{th}$ information bit
in $\boldsymbol{b}_{n}$, and the \emph{a posteriori} probability
$p(b_{n\iota}\mid\boldsymbol{y})$ is given by
\begin{equation}
p\left(b_{n\iota}\mid\boldsymbol{y}\right)\propto\sum_{\boldsymbol{b}\backslash b_{n\iota},\boldsymbol{c},\boldsymbol{x}}\int_{\boldsymbol{H},\boldsymbol{W}}p\left(\boldsymbol{b},\boldsymbol{c},\boldsymbol{x},\boldsymbol{y},\boldsymbol{W},\boldsymbol{H}\right).\label{eq:p_bi_y}
\end{equation}
Since $\boldsymbol{b}\shortrightarrow\boldsymbol{c}\shortrightarrow\boldsymbol{x}\shortrightarrow\boldsymbol{y}$
is a Markov chain and the CFR matrix $\boldsymbol{W}$ only depends
on the CIR matrix $\boldsymbol{H}$, the joint probability $p(\boldsymbol{b},\boldsymbol{c},\boldsymbol{x},\boldsymbol{y},\boldsymbol{W},\boldsymbol{H})$
can be factorized into
\begin{align}
 & p\left(\boldsymbol{b},\boldsymbol{c},\boldsymbol{x},\boldsymbol{y},\boldsymbol{W},\boldsymbol{H}\right)=p\left(\boldsymbol{b}\right)p\left(\boldsymbol{c}\mid\boldsymbol{b}\right)p\left(\boldsymbol{x}\mid\boldsymbol{c}\right)p\left(\boldsymbol{y}\mid\boldsymbol{W},\boldsymbol{x}\right)p\left(\boldsymbol{H},\boldsymbol{W}\right).\label{eq:factorization_p_bcxywh}
\end{align}
The conditional probability $p(\boldsymbol{x}\mid\boldsymbol{c})$
in (\ref{eq:factorization_p_bcxywh}) can be factorized into
\begin{equation}
p\left(\boldsymbol{x}\mid\boldsymbol{c}\right)=\prod_{t}p\left(\boldsymbol{x}_{t}\mid\boldsymbol{c}_{t}\right)=\prod_{t,n,k}p\left(x_{tnk}\mid\boldsymbol{c}_{tnk}\right),
\end{equation}
where $\boldsymbol{c}_{t}\triangleq\{\boldsymbol{c}_{tnk},\forall n,k\}$,
$\boldsymbol{x}_{t}\triangleq\{x_{tnk},\forall n,k\}$, $p(x_{tnk}\mid\boldsymbol{c}_{tnk})=\delta(\varphi(\boldsymbol{c}_{tnk})-x_{tnk})$
denotes the deterministic mapping $x_{tnk}=\varphi(\boldsymbol{c}_{tnk})$,
$\varphi(\boldsymbol{c}_{tnk})$ is the mapping function and $\delta(\cdot)$
is the Kronecker delta function. In practice, the receive correlation
matrices $\{\boldsymbol{R}_{nl}\}$ are unknown, so we impose a conditional
independent structure on the \emph{a priori} probability of $\boldsymbol{H}$,
i.e.,
\begin{align}
p\left(\boldsymbol{H}\mid\boldsymbol{\gamma}\right) & =\prod_{n,l}p\left(\boldsymbol{h}_{\cdot nl}\mid\gamma_{nl}\right),\\
p\left(\boldsymbol{h}_{\cdot nl}\mid\gamma_{nl}\right) & =\prod_{m}p\left(h_{mnl}\mid\gamma_{nl}\right),\\
p\left(h_{mnl}\mid\gamma_{nl}\right) & =\mathcal{N}_{\mathbb{C}}\left(h_{mnl};\gamma_{nl}^{-1}\right),\\
p\left(\gamma_{nl}\right) & =\mathsf{Gam}(\gamma_{nl};0,0),
\end{align}
where $\boldsymbol{\gamma}\triangleq\{\gamma_{nl}\}$, and $\gamma_{nl}$
is the inversion of PDP to be learned. As the CFR $\boldsymbol{w}_{mn\cdot}$
is the Fourier transformations of the CIR $\boldsymbol{h}_{mn\cdot}$,
i.e., $\boldsymbol{w}_{mn\cdot}=\boldsymbol{\varPhi}\boldsymbol{h}_{mn\cdot},\forall m,\forall n,$
then the conditional probability $p(\boldsymbol{W}\mid\boldsymbol{H})$
reads
\begin{align}
p(\boldsymbol{W}\mid\boldsymbol{H}) & =\prod_{m,n}p(\boldsymbol{w}_{mn\cdot}\mid\boldsymbol{h}_{mn\cdot})=\prod_{m,n,k}\delta\left(w_{mnk}-\sum_{l}\phi_{kl}h_{mnl}\right),
\end{align}
where $\boldsymbol{\varPhi}\in\mathbb{C}^{K\times L}$ denotes the
DFT weighting matrix, and $\phi_{kl}$ denotes the entry in the $k\text{th}$
row and $l\text{th}$ column of $\boldsymbol{\varPhi}$. The channel
transition function $p(\boldsymbol{y}\mid\boldsymbol{W},\boldsymbol{x})$
is factorized into
\begin{equation}
p(\boldsymbol{y}\mid\boldsymbol{W},\boldsymbol{x})=\prod_{t,m,k}f_{tmk}\left(\boldsymbol{x}_{t\cdot k},\boldsymbol{w}_{m\cdot k}\right),\label{eq:factorization_channel_transfer_pdf}
\end{equation}
where $\boldsymbol{x}_{t\cdot k}\triangleq[x_{t1k}\cdots x_{tNk}]^{\mathsf{T}}$,
$\boldsymbol{w}_{m\cdot k}\triangleq[w_{m1k}\cdots w_{mNk}]^{\mathsf{T}}$,
and
\begin{equation}
f_{tmk}\left(\boldsymbol{x}_{t\cdot k},\boldsymbol{w}_{m\cdot k}\right)=\mathcal{N}_{\mathbb{C}}\left(y_{tmk};\sum_{n}w_{mnk}x_{mnk},\sigma_{\varpi}^{2}\right).\label{eq:f_=00007Btmk=00007D}
\end{equation}

\begin{figure}
\centering\includegraphics[width=5in]{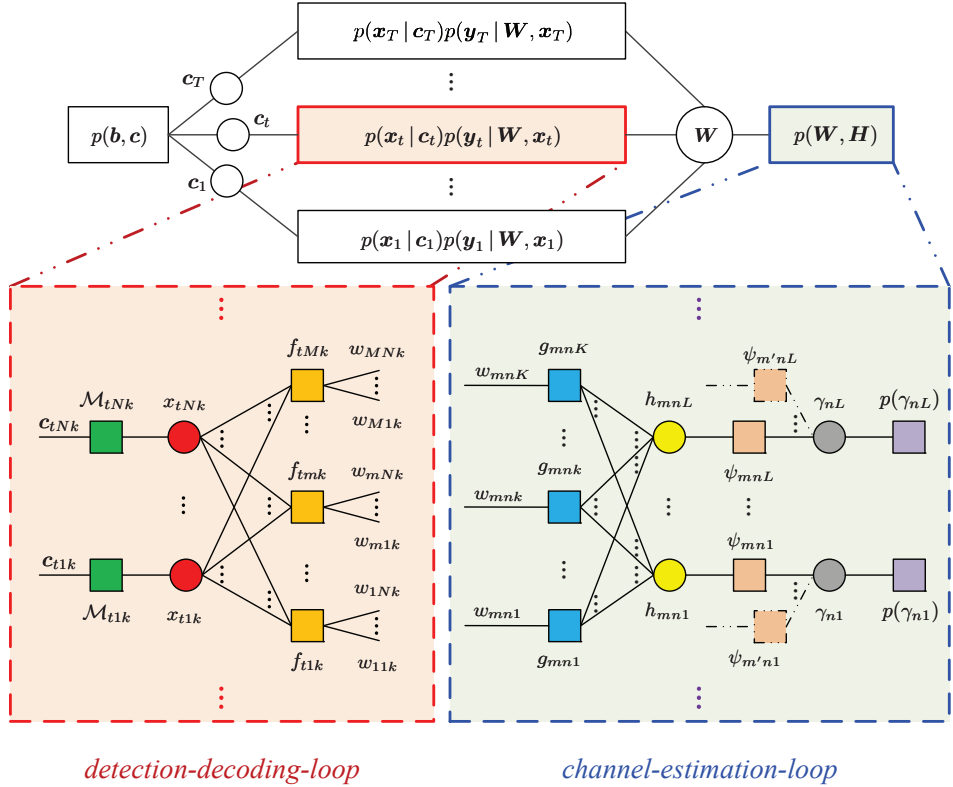}\caption{Factor graph of the massive MIMO-OFDM system.}
\label{fig:factorgraph}
\end{figure}

The probabilistic structure defined by the factorizations (\ref{eq:factorization_p_bcxywh})-(\ref{eq:factorization_channel_transfer_pdf})
can be represented by the factor graph, as depicted in Fig. \ref{fig:factorgraph}.
In this factor graph, mapping constraint $\delta(\varphi(\boldsymbol{c}_{tnk})-x_{tnk})$
appears as a function node $\mathcal{M}_{tnk}$, the mixing constraint
$\delta(w_{mnk}-\sum_{l}\phi_{kl}h_{mnl})$ appears as function node
$g_{mnk}$, and the \emph{a prior }distribution $\psi(h_{mnl},\gamma_{nl})$
appears as function node $\psi_{mnl}$.

 There exist two groups of loops, the \emph{detection-decoding-loop}
on the left and the \emph{channel-estimation-loop} on the right. Unlike
a tree-structured factor graph, the existence of loops implies various
iterative message passing schedules. In our case, we choose to start
passing messages at the channel transition nodes $\{f_{tmk}\}$, then
pass messages concurrently in both the \emph{detection-decoding-loop}
and the \emph{channel-estimation-loop}. Each of these full cycles
of message passing will be referred to as a ``turbo iteration\textquotedblright .

\section{Expectation Propagation for Joint Channel Estimation And Decoding
\label{sec:Hybrid-Message-Passing}}

The presentation of message passing follows closely with the convention
in \cite{Kschischang2001}. Due to high-dimensional integration, directly
applying the SPA to the factor graph in Fig. \ref{fig:factorgraph}
is computationally prohibitive. Hence, we resort to approximate inference
to find efficient solutions.

\subsection{Message Updating in Detection-Decoding-Loop}

Note that, to update the outgoing messages from the channel transition
node $f_{tmk}$, the received signal shown in (\ref{eq:receive signal model})
can be rewritten as
\begin{equation}
y_{tmk}=w_{mnk}x_{tnk}+\sum_{n'\neq n}w_{mn'k}x_{tn'k}+\varpi_{tmk},\forall n.\label{eq:y_tm^k}
\end{equation}
The interference term $\sum_{n'\neq n}w_{mn'k}x_{tn'k}+\varpi_{tmk}$
in (\ref{eq:y_tm^k}) is considered as a Gaussian variable \cite{Som2011,Parker2014Bilinear},
and then $y_{tmk}-\bigl(\sum_{n'\neq n}w_{mn'k}x_{tn'k}+\varpi_{tmk}\bigr)$
is also a Gaussian variable with the mean $z_{f_{tmk}\shortrightarrow x_{tnk}}^{(i)}$
and variance $\tau_{f_{tmk}\shortrightarrow x_{tnk}}^{(i)}$ given
by
\begin{align}
z_{f_{tmk}\shortrightarrow x_{tnk}}^{(i)} & =y_{tmk}-\sum_{n'\neq n}\hat{w}_{w_{mn'k}\shortrightarrow f_{tmk}}^{(i-1)}\hat{x}_{x_{tn'k}\shortrightarrow f_{tmk}}^{(i-1)},\label{eq:hatz_fmk_to_x_nk^i}\\
\tau_{f_{tmk}\shortrightarrow x_{tnk}}^{(i)} & =\sigma_{\varpi}^{2}+\sum_{n'\neq n}\left(\bigl|\hat{w}_{w_{mn'k}\shortrightarrow f_{tmk}}^{(i-1)}\bigr|^{2}\nu_{x_{tn'k}\shortrightarrow f_{tmk}}^{(i-1)}\right.\nonumber \\
 & \left.\hspace{1em}+\nu_{w_{mn'k}\shortrightarrow f_{tmk}}^{(i-1)}\bigl|\hat{x}_{x_{tn'k}\shortrightarrow f_{tmk}}^{(i-1)}\bigr|^{2}+\nu_{w_{mn'k}\shortrightarrow f_{tmk}}^{(i-1)}\nu_{x_{tn'k}\shortrightarrow f_{tmk}}^{(i-1)}\right),\label{eq:tau_fmk_to_x_nk^i-2}
\end{align}
where $\hat{x}_{x_{tnk}\shortrightarrow f_{tmk}}^{(i-1)}$ and $\nu_{x_{tnk}\shortrightarrow f_{tmk}}^{(i-1)}$
denote the mean and variance of variable $x_{tnk}$ with respect to
the message $\mu_{x_{tnk}\shortrightarrow f_{tmk}}^{(i-1)}(x_{tnk})$;
$\hat{w}_{w_{mnk}\shortrightarrow f_{tmk}}^{(i-1)}$ and $\nu_{w_{mnk}\shortrightarrow f_{tmk}}^{(i-1)}$
denote the mean and variance of variable $w_{mnk}$ with respect to
the message $\mu_{w_{mnk}\shortrightarrow f_{tmk}}^{(i-1)}(w_{mnk})$.
From the model shown in (\ref{eq:y_tm^k})-(\ref{eq:tau_fmk_to_x_nk^i-2}),
the channel transition function $f_{tmk}$ at the $i\text{th}$ turbo
iteration can be viewed as
\begin{equation}
\hat{f}_{tmk}^{(i)}(w_{mnk},x_{tnk})=\mathcal{N}_{\mathbb{C}}\left(w_{mnk}x_{tnk};z_{f_{tmk}\shortrightarrow x_{tnk}}^{(i)},\tau_{f_{tmk}\shortrightarrow x_{tnk}}^{(i)}\right),\forall n
\end{equation}
Consequently, the message $\mu_{f_{tmk}\shortrightarrow x_{tnk}}^{(i)}(x_{tnk})$
is calculated by
\begin{align}
\mu_{f_{tmk}\shortrightarrow x_{tnk}}^{(i)}(x_{tnk}) & =\int_{w_{mnk}}\hat{f}_{tmk}^{(i)}(w_{mnk},x_{tnk})\mu_{w_{mnk}\shortrightarrow f_{tmk}}^{(i-1)}(w_{mnk})\nonumber \\
 & \propto\mathcal{N}_{\mathbb{C}}\left(x_{tnk};\frac{z_{f_{tmk}\shortrightarrow x_{tnk}}^{\left(i\right)}}{\hat{w}_{w_{mnk}\shortrightarrow f_{tmk}}^{\left(i-1\right)}},\frac{\tau_{f_{tmk}\shortrightarrow x_{tnk}}^{(i)}+\nu_{w_{mnk}\shortrightarrow f_{tmk}}^{\left(i-1\right)}\left|x_{tnk}\right|^{2}}{\bigl|\hat{w}_{w_{mnk}\shortrightarrow f_{tmk}}^{\left(i-1\right)}\bigl|^{2}}\right).\label{eq:mu_=00007Btnk_shortleftarrow_tmk=00007D}
\end{align}
Using (\ref{eq:mu_=00007Btnk_shortleftarrow_tmk=00007D}), the message
from the variable $x_{tnk}$ to the channel transition node $f_{tmk}$
is updated by
\begin{equation}
\mu_{x_{tnk}\shortrightarrow f_{tmk}}^{(i)}\left(x_{tnk}\right)=\mu_{\mathcal{M}_{tnk}\shortrightarrow x_{tnk}}^{(i)}\left(x_{tnk}\right)\mathsf{exp}\left(-\sum_{m'\neq m}\Delta_{f_{tm'k}\shortrightarrow x_{tnk}}^{(i)}\left(x_{tnk}\right)\right),\label{eq:mu_x_tnk2f_tmk^i}
\end{equation}
where
\begin{equation}
\Delta_{f_{tmk}\shortrightarrow x_{tnk}}^{(i)}\left(x_{tnk}\right)=\frac{\left|z_{f_{tmk}\shortrightarrow x_{tnk}}^{(i)}-\hat{w}_{w_{mnk}\shortrightarrow f_{tmk}}^{\left(i-1\right)}x_{tnk}\right|^{2}}{\tau_{f_{tmk}\shortrightarrow x_{tnk}}^{(i)}+\nu_{w_{mnk}\shortrightarrow f_{tmk}}^{\left(i-1\right)}\left|x_{tnk}\right|^{2}}+\mathsf{ln}\left(\tau_{f_{tmk}\shortrightarrow x_{tnk}}^{(i)}+\nu_{w_{mnk}\shortrightarrow f_{tmk}}^{\left(i-1\right)}\left|x_{tnk}\right|^{2}\right).\label{eq:Delta_=00007Bf_=00007Btmk=00007Dx_=00007Btnk=00007D=00007D}
\end{equation}
To obtain $z_{f_{tmk}\shortrightarrow x_{tnk}}^{(i)}$ in (\ref{eq:hatz_fmk_to_x_nk^i})
and $\tau_{f_{tmk}\shortrightarrow x_{tnk}}^{(i)}$ in (\ref{eq:tau_fmk_to_x_nk^i-2}),
the mean and variance of variable $x_{tnk}$ with respect to the message
$\mu_{x_{tnk}\shortrightarrow f_{tmk}}^{(i-1)}(x_{tnk})$ are calculated
by
\begin{align}
 & \hat{x}_{x_{tnk}\shortrightarrow f_{tmk}}^{\left(i-1\right)}=\frac{\sum_{\alpha_{s}\in\mathcal{A}}\alpha_{s}\mu_{x_{tnk}\shortrightarrow f_{tmk}}^{(i-1)}\left(x_{tnk}=\alpha_{s}\right)}{\sum_{\alpha_{s}\in\mathcal{A}}\mu_{x_{tnk}\shortrightarrow f_{tmk}}^{(i-1)}\left(x_{tnk}=\alpha_{s}\right)},\label{eq:hatx_xnk_to_fmk^i}\\
 & \nu_{x_{tnk}\shortrightarrow f_{tmk}}^{\left(i-1\right)}=\frac{\sum_{\alpha_{s}\in\mathcal{A}}\left|\alpha_{s}\right|^{2}\mu_{x_{tnk}\shortrightarrow f_{tmk}}^{(i-1)}\left(x_{tnk}=\alpha_{s}\right)}{\sum_{\alpha_{s}\in\mathcal{A}}\mu_{x_{tnk}\shortrightarrow f_{tmk}}^{(i-1)}\left(x_{tnk}=\alpha_{s}\right)}-\bigl|\hat{x}_{x_{tnk}\shortrightarrow f_{tmk}}^{\left(i-1\right)}\bigr|^{2}.\label{eq:nu_xnk_to_fmk^i}
\end{align}
Using the Gaussian approximation shown in (\ref{eq:y_tm^k})-(\ref{eq:tau_fmk_to_x_nk^i-2})
again, the message $\mu_{f_{tmk}\shortrightarrow w_{mnk}}^{(i)}\left(w_{mnk}\right)$
is then updated by
\begin{equation}
\mu_{f_{tmk}\shortrightarrow w_{mnk}}^{(i)}\left(w_{mnk}\right)\propto\sum_{x_{tnk}\in\mathcal{A}}\vartheta_{f_{tmk}}^{(i)}\left(x_{tnk}\right)\mathcal{N}_{\mathbb{C}}\left(w_{mnk};\frac{z_{f_{tmk}\shortrightarrow x_{tnk}}^{(i)}}{x_{tnk}},\frac{\tau_{f_{tmk}\shortrightarrow x_{tnk}}^{(i)}}{\left|x_{tnk}\right|^{2}}\right),\label{eq:mu_tmk2mnk}
\end{equation}
where $\vartheta_{f_{tmk}}^{(i)}\left(x_{tnk}\right)$ denotes the
weight of Gaussian component,
\begin{equation}
\vartheta_{f_{tmk}}^{(i)}\left(x_{tnk}\right)=\frac{\left|x_{tnk}\right|^{-2}\mu_{x_{tnk}\shortrightarrow f_{tmk}}^{(i-1)}\left(x_{tnk}\right)}{\sum_{x_{tnk}\in\mathcal{A}}\left|x_{tnk}\right|^{-2}\mu_{x_{tnk}\shortrightarrow f_{tmk}}^{(i-1)}\left(x_{tnk}\right)},x_{tnk}\in\mathcal{A}.
\end{equation}
 As $\mu_{f_{tmk}\shortrightarrow w_{mnk}}^{(i)}\left(w_{mnk}\right)$
given by (\ref{eq:mu_tmk2mnk}) is a Gaussian mixture, the number
of its components will increase exponentially in the consequent message
updating. To avoid the increase, the message $\mu_{f_{tmk}\shortrightarrow w_{mnk}}^{(i)}(w_{mnk})$
can be projected onto a Gaussian function by the criterion of minimum
KL divergence as in \cite{novak2013idma} and \cite{liu2009joint}.
The projection reduces to matching the first two order moments of
a Gaussian function $\mathcal{N}_{\mathbb{C}}(w_{mnk};\hat{w}_{f_{tmk}\shortrightarrow w_{mnk}}^{(i)},\nu_{f_{tmk}\shortrightarrow w_{mnk}}^{(i)})$
and the message $\mu_{f_{tmk}\shortrightarrow w_{mnk}}^{(i)}(w_{mnk})$
\cite{bishop2006pattern}, leading to
\begin{align}
\hat{w}_{f_{tmk}\shortrightarrow w_{mnk}}^{(i)} & =z_{f_{tmk}\shortrightarrow x_{tnk}}^{(i)}\sum_{x_{tnk}\in\mathcal{A}}\frac{\vartheta_{f_{tmk}}^{(i)}\left(x_{tnk}\right)}{x_{tnk}},\label{eq:v_fjk2gjik-1-1}\\
\nu_{f_{tmk}\shortrightarrow w_{mnk}}^{(i)} & =\left(\tau_{f_{tmk}\shortrightarrow x_{tnk}}^{(i)}+\left|z_{f_{tmk}\shortrightarrow x_{tnk}}^{(i)}\right|^{2}\right)\sum_{x_{tnk}\in\mathcal{A}}\frac{\vartheta_{f_{tmk}}^{(i)}(x_{tnk})}{\left|x_{tnk}\right|^{2}}-\left|\hat{w}_{f_{tmk}\shortrightarrow w_{mnk}}^{(i)}\right|^{2}.\label{eq:eq:m_fjk2gjik-1-1}
\end{align}
The Gaussian approximations shown in (\ref{eq:y_tm^k})-(\ref{eq:eq:m_fjk2gjik-1-1})
lead to a desirable closed-form message passing algorithm, which will
be referred to as ``BP-GA''. However, it bears a heavy computations
burden: it needs to calculate each $\mu_{x_{tnk}\shortrightarrow f_{tmk}}^{(i)}(x_{tnk}),\forall x_{tnk}\in\mathcal{A}$,
but the term $-\sum_{m'\neq m}\Delta_{f_{tm'k}\shortrightarrow x_{tnk}}^{(i)}(x_{tnk})$
in (\ref{eq:mu_x_tnk2f_tmk^i}) is complex as $M$ is large in the
massive MIMO systems. Besides, it needs to calculate each $\hat{x}_{x_{tnk}\shortrightarrow f_{tmk}}^{(i-1)}$
and $\nu_{x_{tnk}\shortrightarrow f_{tmk}}^{(i-1)}$ using (\ref{eq:hatx_xnk_to_fmk^i})
and (\ref{eq:nu_xnk_to_fmk^i}), which amounts to $TMNK$.

Next, we will derive an efficient message-passing algorithm by the
framework of expectation propagation. Recalling (\ref{eq:y_tm^k})-(\ref{eq:tau_fmk_to_x_nk^i-2}),
a local belief of $w_{mnk}$ at the channel-transition function $f_{tmk}$
can be defined by
\begin{align}
\beta_{f_{tmk}}^{(i)}\left(w_{mnk}\right) & =\mu_{w_{mnk}\shortrightarrow f_{tmk}}^{(i-1)}\left(w_{mnk}\right)\int_{x_{tnk}}\hat{f}_{tmk}^{(i)}(w_{mnk},x_{tnk})\mu_{x_{tnk}\shortrightarrow f_{tmk}}^{(i-1)}\left(x_{tnk}\right)\nonumber \\
 & \propto\mathsf{exp}\left(-\Delta_{f_{tmk}}^{(i)}\left(w_{mnk}\right)\right),\forall n,\label{eq:beta_fmnk^i_xnk}
\end{align}
where
\begin{align}
\Delta_{f_{mnk}}^{(i)}(w_{mnk}) & =\frac{\left|z_{f_{tmk}\shortrightarrow x_{tnk}}^{(i)}-\hat{x}_{x_{tnk}\shortrightarrow f_{tmk}}^{(i-1)}w_{mnk}\right|^{2}}{\tau_{f_{tmk}\shortrightarrow x_{tnk}}^{(i)}+\nu_{x_{tnk}\shortrightarrow f_{tmk}}^{(i-1)}\left|w_{mnk}\right|^{2}}+\frac{\left|w_{mnk}-w_{w_{mnk}\shortrightarrow f_{tmk}}^{(i-1)}\right|^{2}}{\nu_{w_{mnk}\shortrightarrow f_{tmk}}^{(i-1)}}\nonumber \\
 & \hspace{1em}+\mathsf{ln}\left(\tau_{f_{tmk}\shortrightarrow x_{tnk}}^{(i)}+\nu_{x_{tnk}\shortrightarrow f_{tmk}}^{(i-1)}\left|w_{mnk}\right|^{2}\right).\label{eq:Delta_fmnk^i_xnk}
\end{align}
We impose a continuous complex Gaussian distribution constraint on
the belief of $w_{mnk}$, i.e., we project $\beta_{f_{tmk}}^{(i)}(w_{mnk})$
to a Gaussian distribution. The projection reduces to a moment matching;
however, the mean and variance of $\beta_{f_{tmk}}^{(i)}(w_{mnk})$
involve complex integrals and there are no analytical solutions. So
we resort to quadratic approximation for calculating the first two
moments of $\beta_{f_{tmk}}^{(i)}(w_{mnk})$.

The term $z_{f_{tmk}\shortrightarrow x_{tnk}}^{(i)}-\hat{x}_{x_{tnk}\shortrightarrow f_{tmk}}^{(i-1)}w_{mnk}$,
$\tau_{f_{tmk}\shortrightarrow x_{tnk}}^{(i)}+\nu_{x_{tnk}\shortrightarrow f_{tmk}}^{(i-1)}\left|w_{mnk}\right|^{2}$,
and $w_{mnk}-w_{w_{mnk}\shortrightarrow f_{tmk}}^{(i-1)}$ in (\ref{eq:Delta_fmnk^i_xnk})
can be rewritten as
\begin{align}
z_{f_{tmk}\shortrightarrow x_{tnk}}^{(i)}-\hat{x}_{x_{tnk}\shortrightarrow f_{tmk}}^{(i-1)}w_{mnk} & =\underset{\hat{z}_{f_{tmk}\shortrightarrow w_{mnk}}^{(i)}}{\underbrace{z_{f_{tmk}\shortrightarrow x_{tnk}}^{(i)}-\hat{x}_{x_{tnk}\shortrightarrow f_{tmk}}^{(i-1)}\hat{w}_{w_{mnk}}^{(i-1)}}}\nonumber \\
 & \hspace{1em}+\hat{x}_{x_{tnk}\shortrightarrow f_{tmk}}^{(i-1)}\bigl(\hat{w}_{w_{mnk}}^{(i-1)}-w_{mnk}\bigr),\\
\tau_{f_{tmk}\shortrightarrow x_{tnk}}^{(i)}+\nu_{x_{tnk}\shortrightarrow f_{tmk}}^{(i-1)}\left|w_{mnk}\right|^{2} & =\underset{\hat{\tau}_{f_{tmk}\shortrightarrow w_{mnk}}^{(i)}}{\underbrace{\tau_{f_{tmk}\shortrightarrow x_{tnk}}^{(i)}+\nu_{x_{tnk}\shortrightarrow f_{tmk}}^{(i-1)}\left|\hat{w}_{w_{mnk}}^{(i-1)}\right|^{2}}}\nonumber \\
 & \hspace{1em}+\nu_{x_{tnk}\shortrightarrow f_{tmk}}^{(i-1)}\left(\left|w_{mnk}\right|^{2}-\left|\hat{w}_{w_{mnk}}^{(i-1)}\right|^{2}\right),\\
w_{mnk}-\hat{w}_{w_{mnk}\shortrightarrow f_{tmk}}^{(i-1)} & =\hat{w}_{w_{mnk}}^{(i-1)}-\hat{w}_{w_{mnk}\shortrightarrow f_{tmk}}^{(i-1)}+\left(w_{mnk}-\hat{w}_{w_{mnk}}^{(i-1)}\right).
\end{align}
where $\hat{w}_{w_{mnk}}^{(i-1)}$ is the \emph{a posteriori} mean
of $w_{mnk}$ at previous turbo iteration. By (\ref{eq:=00007BH=00007D(=00007B=00007Bu=00007D=00007D,=00007B=00007Bv=00007D=00007D)})
shown in the Appendix, we can expand $\Delta_{f_{tmk}}^{(i)}(x_{tnk})$
at the point $\vec{\boldsymbol{z}}_{0}=[\hat{z}_{f_{tmk}\shortrightarrow w_{mnk}}^{(i)},(\hat{z}_{f_{tmk}\shortrightarrow w_{mnk}}^{(i)})^{*}]$,
$\tau_{0}=\hat{\tau}_{f_{tmk}\shortrightarrow w_{mnk}}^{(i)}$, and
$\vec{\boldsymbol{u}}_{0}=[\hat{w}_{w_{mnk}}^{(i-1)}-\hat{w}_{w_{mnk}\shortrightarrow f_{tmk}}^{(i-1)},(\hat{w}_{w_{mnk}}^{(i-1)}-\hat{w}_{w_{mnk}\shortrightarrow f_{tmk}}^{(i-1)})^{*}]$,
i.e.,
\begin{align}
\Delta_{f_{mnk}}^{(i)}(w_{mnk}) & =\Biggl(\frac{1}{\nu_{w_{mnk}\shortrightarrow f_{tmk}}^{(i-1)}}+\frac{\bigl|\hat{x}_{x_{tnk}\shortrightarrow f_{tmk}}^{(i-1)}\bigr|^{2}}{\hat{\tau}_{f_{tmk}\shortrightarrow w_{mnk}}^{(i)}}+\frac{\nu_{x_{tnk}\shortrightarrow f_{tmk}}^{(i-1)}}{\hat{\tau}_{f_{tmk}\shortrightarrow w_{mnk}}^{(i)}}\biggl(1-\frac{\hat{z}_{f_{tmk}\shortrightarrow w_{mnk}}^{(i)}\bigr|^{2}}{\hat{\tau}_{f_{tmk}\shortrightarrow w_{mnk}}^{(i)}}\biggr)\Biggr)\left|w_{mnk}\right|^{2}\nonumber \\
 & \hspace{1em}-2\Re\left\{ \frac{\bigl(z_{f_{tmk}\shortrightarrow x_{tnk}}^{(i)}\bigr)^{*}\hat{x}_{x_{tnk}\shortrightarrow f_{tmk}}^{(i-1)}}{\hat{\tau}_{f_{tmk}\shortrightarrow w_{mnk}}^{(i)}}w_{mnk}+\frac{\bigl(\hat{w}_{w_{mnk}\shortrightarrow f_{tmk}}^{(i-1)}\bigr)^{*}}{\nu_{w_{mnk}\shortrightarrow f_{tmk}}^{(i-1)}}w_{mnk}\right\} +\mathsf{const},\label{eq:Delta_=00007Bf_=00007Bmn=00007D^=00007Bk=00007D=00007D^=00007B(i)=00007D(w_=00007Bmnk=00007D)}
\end{align}
where the invariant terms with respect to $w_{mnk}$ are absorbed
into the constant term $\mathsf{const}$. Note that using (\ref{eq:Delta_=00007Bf_=00007Bmn=00007D^=00007Bk=00007D=00007D^=00007B(i)=00007D(w_=00007Bmnk=00007D)})
$\mathsf{exp}\{-\Delta_{f_{mn}^{k}}^{(i)}(w_{mnk})\}$ is essentially
the Gaussian approximation of $\beta_{f_{tmk}}^{(i)}(w_{mnk})$, i.e.,
\begin{equation}
\beta_{f_{tmk}}^{(i)}(w_{mnk})\approx\mathcal{N}_{\mathbb{C}}\left(w_{mnk};\hat{w}_{f_{tmk}}^{(i)},\nu_{f_{tmk}}^{(i)}\right),\label{eq:beta_=00007Bf_=00007Btmk=00007D=00007D^=00007B(i)=00007D(w_=00007Bmnk=00007D)}
\end{equation}
where
\begin{align}
 & \nu_{f_{tmk}}^{(i)}=\frac{\hat{\tau}_{f_{tmk}\shortrightarrow w_{mnk}}^{(i)}}{\frac{\hat{\tau}_{f_{tmk}\shortrightarrow w_{mnk}}^{(i)}}{\nu_{w_{mnk}\shortrightarrow f_{tmk}}^{(i-1)}}+\bigl|\hat{x}_{x_{tnk}\shortrightarrow f_{tmk}}^{(i-1)}\bigr|^{2}+\nu_{x_{tnk}\shortrightarrow f_{tmk}}^{(i-1)}\biggl(1-\frac{\hat{z}_{f_{tmk}\shortrightarrow w_{mnk}}^{(i)}\bigr|^{2}}{\hat{\tau}_{f_{tmk}\shortrightarrow w_{mnk}}^{(i)}}\biggr)},\\
 & \hat{w}_{f_{tmk}}^{(i)}=\nu_{f_{tmk}}^{(i)}\left(\frac{\left(\hat{x}_{x_{tnk}\shortrightarrow f_{tmk}}^{(i-1)}\right)^{*}z_{f_{tmk}\shortrightarrow x_{tnk}}^{(i)}}{\hat{\tau}_{f_{tmk}\shortrightarrow w_{mnk}}^{(i)}}+\frac{\hat{w}_{w_{mnk}\shortrightarrow f_{tmk}}^{(i-1)}}{\nu_{w_{mnk}\shortrightarrow f_{tmk}}^{(i-1)}}\right).
\end{align}
Using the expectation propagation principle and (\ref{eq:beta_=00007Bf_=00007Btmk=00007D=00007D^=00007B(i)=00007D(w_=00007Bmnk=00007D)}),
we get
\begin{equation}
\mu_{f_{tmk}\shortrightarrow w_{mnk}}^{(i)}=\frac{\beta_{f_{tmk}}^{(i)}\left(w_{mnk}\right)}{\mu_{w_{mnk}\shortrightarrow f_{tmk}}^{(i-1)}}\propto\mathcal{N}_{\mathbb{C}}\left(w_{mnk};\hat{w}_{f_{tmk}\shortrightarrow w_{mnk}}^{(i)},\nu_{f_{tmk}\shortrightarrow w_{mnk}}^{(i)}\right),
\end{equation}
where
\begin{align}
\nu_{f_{tmk}\shortrightarrow w_{mnk}}^{(i)} & =\frac{\hat{\tau}_{f_{tmk}\shortrightarrow w_{mnk}}^{(i)}}{\bigl|\hat{x}_{x_{tnk}\shortrightarrow f_{tmk}}^{(i-1)}\bigr|^{2}+\nu_{x_{tnk}\shortrightarrow f_{tmk}}^{(i-1)}\biggl(1-\frac{\hat{z}_{f_{tmk}\shortrightarrow w_{mnk}}^{(i)}\bigr|^{2}}{\hat{\tau}_{f_{tmk}\shortrightarrow w_{mnk}}^{(i)}}\biggr)},\label{eq:nu_=00007Bf_=00007Btmk=00007D2w_=00007Bmnk=00007D=00007D^=00007B(i)=00007D}\\
\hat{w}_{f_{tmk}\shortrightarrow w_{mnk}}^{(i)} & =\nu_{f_{tmk}\shortrightarrow w_{mnk}}^{(i)}\frac{\left(\hat{x}_{x_{tnk}\shortrightarrow f_{tmk}}^{(i-1)}\right)^{*}z_{f_{tmk}\shortrightarrow x_{tnk}}^{(i)}}{\hat{\tau}_{f_{tmk}\shortrightarrow w_{mnk}}^{(i)}}.
\end{align}
Note that, the messages at the channel transition nodes associated
with known pilot symbol boil down to the following simple form
\begin{equation}
\mu_{f_{tmk}\shortrightarrow w_{mnk}}^{(i)}(w_{mnk})\propto\mathcal{N}_{\mathbb{C}}\left(w_{mnk};\frac{y_{tmk}}{x_{tnk}},\frac{\sigma_{\varpi}^{2}}{\left|x_{tnk}\right|^{2}}\right),\forall(t,k)\in\mathcal{P}_{n},
\end{equation}
where we use the fact that other users transmit zero-symbols on the
pilot subcarriers $\mathcal{P}_{n}$, and pilot symbol $x_{tnk}$
takes a known value.

Similarly, at the channel-transition function $f_{tmk}$, a local
belief of $x_{tnk}$ can be defined by
\begin{equation}
\beta_{f_{mnk}}^{(i)}(x_{tnk})\propto\mathsf{exp}\left(-\Delta_{f_{tmk}}^{(i)}\left(x_{tnk}\right)\right),
\end{equation}
where
\begin{equation}
\Delta_{f_{tmk}}^{(i)}\left(x_{tnk}\right)=\Delta_{f_{tmk}\shortrightarrow x_{tnk}}^{(i)}\left(x_{tnk}\right)+\frac{\left|x_{tnk}-\hat{x}{}_{x_{tnk}\shortrightarrow f_{tmk}}^{(i-1)}\right|^{2}}{\nu_{x_{tnk}\shortrightarrow f_{tmk}}^{(i-1)}}.\label{eq:Delta_=00007Bf_=00007Btmk=00007D=00007D^=00007B(i)=00007D(x_=00007Btnk=00007D)}
\end{equation}
The term $z_{f_{tmk}\shortrightarrow x_{tnk}}^{(i)}-\hat{w}_{w_{mnk}\shortrightarrow f_{tmk}}^{\left(i-1\right)}x_{tnk}$,
$\tau_{f_{tmk}\shortrightarrow x_{tnk}}^{(i)}+\nu_{w_{mnk}\shortrightarrow f_{tmk}}^{\left(i-1\right)}\left|x_{tnk}\right|^{2}$
and $x_{tnk}-\hat{x}{}_{x_{tnk}\shortrightarrow f_{tmk}}^{(i-1)}$
in (\ref{eq:Delta_=00007Bf_=00007Btmk=00007D=00007D^=00007B(i)=00007D(x_=00007Btnk=00007D)})
can also be rewritten as
\begin{align}
z_{f_{tmk}\shortrightarrow x_{tnk}}^{(i)}-\hat{w}_{w_{mnk}\shortrightarrow f_{tmk}}^{\left(i-1\right)}x_{tnk} & =\underset{\hat{z}_{f_{tmk}\shortrightarrow x_{tnk}}^{(i)}}{\underbrace{z_{f_{tmk}\shortrightarrow x_{tnk}}^{(i)}-\hat{w}_{w_{mnk}\shortrightarrow f_{tmk}}^{\left(i-1\right)}\hat{x}_{x_{tnk}}^{(i-1)}}}\nonumber \\
 & \hspace{1em}+\hat{w}_{w_{mnk}\shortrightarrow f_{tmk}}^{\left(i-1\right)}\left(\hat{x}_{x_{tnk}}^{(i-1)}-x_{tnk}\right),\\
\tau_{f_{tmk}\shortrightarrow x_{tnk}}^{(i)}+\nu_{w_{mnk}\shortrightarrow f_{tmk}}^{\left(i-1\right)}\left|x_{tnk}\right|^{2} & =\underset{\hat{\tau}_{f_{tmk}\shortrightarrow x_{tnk}}^{(i)}}{\underbrace{\tau_{f_{tmk}\shortrightarrow x_{tnk}}^{(i)}+\nu_{w_{mnk}\shortrightarrow f_{tmk}}^{(i-1)}\left|\hat{x}_{x_{tnk}}^{(i-1)}\right|^{2}}}\nonumber \\
 & \hspace{1em}+\nu_{w_{mnk}\shortrightarrow f_{tmk}}^{\left(i-1\right)}\left(\left|x_{tnk}\right|^{2}-\left|\hat{x}_{x_{tnk}}^{(i-1)}\right|^{2}\right),\\
x_{tnk}-\hat{x}{}_{x_{tnk}\shortrightarrow f_{tmk}}^{(i-1)} & =\hat{x}_{x_{tnk}}^{(i-1)}-\hat{x}{}_{x_{tnk}\shortrightarrow f_{tmk}}^{(i-1)}+\left(x_{tnk}-\hat{x}_{x_{tnk}}^{(i-1)}\right),
\end{align}
where $\hat{x}_{x_{tnk}}^{(i-1)}$ is the \emph{a posteriori} mean
of $x_{tnk}$ at previous turbo iteration. Then $\Delta_{f_{tmk}}^{(i)}(x_{tnk})$
is expanded at the point at the point $\vec{\boldsymbol{z}}_{0}=[\hat{z}_{f_{tmk}\shortrightarrow x_{tnk}}^{(i)},(\hat{z}_{f_{tmk}\shortrightarrow x_{tnk}}^{(i)})^{*}]$,
$\tau_{0}=\hat{\tau}_{f_{tmk}\shortrightarrow x_{tnk}}^{(i)}$, and
$\vec{\boldsymbol{u}}_{0}=[\hat{x}_{x_{tnk}}^{(i-1)}-\hat{x}{}_{x_{tnk}\shortrightarrow f_{tmk}}^{(i-1)},(\hat{x}_{x_{tnk}}^{(i-1)}-\hat{x}{}_{x_{tnk}\shortrightarrow f_{tmk}}^{(i-1)})^{*}]$,
and we have
\begin{equation}
\mu_{f_{tmk}\shortrightarrow x_{tnk}}^{(i)}\left(x_{tnk}\right)\propto\mathcal{N}_{\mathbb{C}}\left(x_{tnk};\hat{x}_{f_{tmk}\shortrightarrow x_{tnk}}^{(i)},\nu_{f_{tmk}\shortrightarrow x_{tnk}}^{(i)}\right),\label{eq:mu_ftmk2w_mnk^i_qa}
\end{equation}
where $\hat{x}_{f_{tmk}\shortrightarrow x_{tnk}}^{(i)}$ and $\nu_{f_{tmk}\shortrightarrow x_{tnk}}^{(i)}$
are given by
\begin{align}
\nu_{f_{tmk}\shortrightarrow x_{tnk}}^{(i)} & =\frac{\hat{\tau}_{f_{tmk}\shortrightarrow x_{tnk}}^{(i)}}{\bigl|\hat{w}_{w_{mnk}\shortrightarrow f_{tmk}}^{(i-1)}\bigr|^{2}+\nu_{w_{mnk}\shortrightarrow f_{tmk}}^{(i-1)}\biggl(1-\frac{\left|\hat{z}_{f_{tmk}\shortrightarrow x_{tnk}}^{(i)}\right|^{2}}{\hat{\tau}_{f_{tmk}\shortrightarrow x_{tnk}}^{(i)}}\biggr)},\\
\hat{x}_{f_{tmk}\shortrightarrow x_{tnk}}^{(i)} & =\nu_{f_{tmk}\shortrightarrow x_{tnk}}^{(i)}\frac{\bigl(\hat{w}_{w_{mnk}\shortrightarrow f_{tmk}}^{(i-1)}\bigr)^{*}z_{f_{tmk}\shortrightarrow x_{tnk}}^{(i)}}{\hat{\tau}_{f_{tmk}\shortrightarrow x_{tnk}}^{(i)}}.
\end{align}

The message $\mu_{x_{tnk}\shortrightarrow\mathcal{M}_{tnk}}^{(i)}(x_{tnk})$
from the variable node $x_{tnk}$ to the mapper node $\mathcal{M}_{tnk}$
is updated by
\begin{equation}
\mu_{x_{tnk}\shortrightarrow\mathcal{M}_{tnk}}^{(i)}\left(x_{tnk}\right)=\prod_{m}\mu_{f_{tmk}\shortrightarrow x_{tnk}}^{(i)}\left(x_{tnk}\right)\propto\mathcal{N}_{\mathbb{C}}\left(x_{tnk};\zeta_{x_{tnk}}^{(i)},\gamma_{x_{tnk}}^{(i)}\right),
\end{equation}
where $\gamma_{x_{tnk}}^{(i)}=1/\sum_{m}\bigl(1/\nu_{f_{tmk}\shortrightarrow x_{tnk}}^{(i)}\bigr)$
and $\zeta_{x_{tnk}}^{(i)}=\gamma_{x_{tnk}}^{(i)}\sum_{m}\bigl(\hat{x}_{f_{tmk}\shortrightarrow x_{tnk}}^{(i)}/\nu_{f_{tmk}\shortrightarrow x_{tnk}}^{(i)}\bigr)$.
With the message $\mu_{x_{tnk}\shortrightarrow\mathcal{M}_{tnk}}^{(i)}(x_{tn}^{k})$
and the\emph{ a priori} LLRs $\{\lambda_{a}^{(i)}(c_{tnk}^{q}),\forall q\}$
fed back by the decoder of user $n$ at the previous turbo iteration,
the extrinsic LLRs $\{\lambda_{e}^{(i)}(c_{tnk}^{q}),\forall q\}$
corresponding to the symbol $x_{tnk}$ are mapped by
\begin{align}
\lambda_{e}^{(i)}\left(c_{tnk}^{q}\right) & =\mathsf{ln}\frac{\sum_{x_{tnk}\in\mathcal{A}_{q}^{1}}\mu_{x_{tnk}\shortrightarrow\mathcal{M}_{tnk}}^{(i)}(x_{tnk})\mu_{\mathcal{M}_{tnk}\shortrightarrow x_{tnk}}^{(i-1)}(x_{tnk})}{\sum_{x_{tnk}\in\mathcal{A}_{q}^{0}}\mu_{x_{tnk}\shortrightarrow\mathcal{M}_{tnk}}^{(i)}(x_{tnk})\mu_{\mathcal{M}_{tnk}\shortrightarrow x_{tnk}}^{(i-1)}(x_{tnk})}-\lambda_{a}^{(i-1)}(c_{tnk}^{q}),\label{eq:lambda_=00007Be=00007D^=00007B(i)=00007D(c_=00007Btnk=00007D^=00007Bq=00007D)}
\end{align}
where the $(i-1)\text{th}$ message $\mu_{\mathcal{M}_{tnk}\shortrightarrow x_{tnk}}^{(i-1)}(x_{tnk})$
is given in the following by (\ref{eq:mu_tnk_shortrightarrow_tnk}).
Once the extrinsic LLRs $\{\lambda_{e}^{(i)}(c_{tnk}^{q})\}$ are
available, each channel decoder performs decoding and feeds back the\emph{
a priori} LLRs of coded bits $\{\lambda_{a}^{(i)}(c_{tnk}^{q})\}$,
which then are interleaved and converted to the following message
\begin{equation}
\mu_{\mathcal{M}_{tnk}\shortrightarrow x_{tnk}}^{(i)}\left(x_{tnk}\right)=\prod_{q}\frac{\mathsf{exp}\left(c_{tnk}^{q}\cdot\lambda_{a}^{(i)}\left(c_{tnk}^{q}\right)\right)}{1+\mathsf{exp}\left(\lambda_{a}^{(i)}\left(c_{tnk}^{q}\right)\right)}.\label{eq:mu_tnk_shortrightarrow_tnk}
\end{equation}

At the variable nodes $\{x_{tnk}\}$, the number of message parameters
$\{\hat{x}_{x_{tnk}\shortrightarrow f_{tmk}}^{\left(i\right)},\nu_{x_{tnk}\shortrightarrow f_{tmk}}^{\left(i\right)}\}$
reaches up to $2TMNK$, so directly evaluating them is expensive via
moment matching like (\ref{eq:hatx_xnk_to_fmk^i}) and (\ref{eq:nu_xnk_to_fmk^i}).
Following the expectation propagation method, we can reduce the computational
complexity of $\{\hat{x}_{x_{tnk}\shortrightarrow f_{tmk}}^{\left(i\right)},\nu_{x_{tnk}\shortrightarrow f_{tmk}}^{\left(i\right)}\}$.
First, at the variable node $x_{tnk}$, the local belief of $x_{tnk}$
is defined by
\begin{equation}
\beta_{x_{tnk}}^{\left(i\right)}(x_{tnk})=\frac{\mu_{\mathcal{M}_{tnk}\shortrightarrow x_{tnk}}^{(i)}(x_{tnk})\mu_{x_{tnk}\shortrightarrow\mathcal{M}_{tnk}}^{(i)}(x_{tnk})}{\sum_{x_{tnk}\in\mathcal{A}}\mu_{\mathcal{M}_{tnk}\shortrightarrow x_{tnk}}^{(i)}(x_{tnk})\mu_{x_{tnk}\shortrightarrow\mathcal{M}_{tnk}}^{(i)}(x_{tnk})}.
\end{equation}
The local belief $\beta_{x_{tnk}}^{\left(i\right)}(x_{tnk})$ can
be projected onto a Gaussian PDF denoted by $\hat{\beta}_{x_{tnk}}^{\left(i\right)}(x_{tnk})=\mathcal{N}_{\mathbb{C}}\bigl(x_{tnk};\hat{x}_{x_{tnk}}^{\left(i\right)},\nu_{x_{tnk}}^{(i)}\bigr)$,
where
\begin{align}
\hat{x}_{x_{tnk}}^{\left(i\right)} & =\sum_{\alpha_{s}\in\mathcal{A}}\alpha_{s}\beta_{x_{tnk}}^{\left(i\right)}\left(x_{tnk}=\alpha_{s}\right),\\
\nu_{x_{tnk}}^{(i)} & =\sum_{\alpha_{s}\in\mathcal{A}}\left|\alpha_{s}\right|^{2}\beta_{x_{tnk}}^{\left(i\right)}\left(x_{tnk}=\alpha_{s}\right)-\left|\hat{x}_{x_{tnk}}^{\left(i\right)}\right|^{2},\label{eq:nu_=00007Bx_=00007Btnk=00007D=00007D^=00007B(i)=00007D}
\end{align}
and then the message $\mu_{x_{tnk}\shortrightarrow f_{tmk}}^{(i)}(x_{tnk})$
is approximated by\cite{Hu2006}
\begin{align}
\hat{\mu}_{x_{tnk}\shortrightarrow f_{tmk}}^{(i)}\left(x_{tnk}\right) & \approx\frac{\hat{\beta}_{x_{tnk}}^{\left(i\right)}\left(x_{tnk}\right)}{\mu_{f_{tmk}\shortrightarrow x_{tnk}}^{(i)}\left(x_{tnk}\right)}\propto\mathcal{N}_{\mathbb{C}}\left(x_{tnk};\hat{x}_{x_{tnk}\shortrightarrow f_{tmk}}^{\left(i\right)},\nu_{x_{tnk}\shortrightarrow f_{tmk}}^{\left(i\right)}\right),\label{eq:tilde_mu_x_i}
\end{align}
where
\begin{align}
\nu_{x_{tnk}\shortrightarrow f_{tmk}}^{(i)} & =\nu_{x_{tnk}}^{(i)}\frac{\nu_{f_{tmk}\shortrightarrow x_{tnk}}^{(i)}}{\nu_{f_{tmk}\shortrightarrow x_{tnk}}^{(i)}-\nu_{x_{tnk}}^{(i)}},\label{eq:nu_=00007Bx_=00007Btnk=00007D2f_=00007Btmk=00007D=00007D^=00007B(i)=00007D}\\
\hat{x}_{x_{tnk}\shortrightarrow f_{tmk}}^{(i)} & =\hat{x}_{x_{tnk}}^{\left(i\right)}+\nu_{x_{tnk}}^{(i)}\frac{\hat{x}_{x_{tnk}}^{\left(i\right)}-\hat{x}_{f_{tmk}\shortrightarrow x_{tnk}}^{(i)}}{\nu_{f_{tmk}\shortrightarrow x_{tnk}}^{(i)}-\nu_{x_{tnk}}^{(i)}}.
\end{align}
\begin{table}
\rule{1\linewidth}{0.5pt}

\begin{align*}
{\scriptstyle \forall t,m,k,n:} & {\scriptstyle z_{f_{tmk}\shortrightarrow x_{tnk}}^{(i)}=y_{tmk}-\sum_{n'\neq n}\hat{w}_{w_{mn'k}\shortrightarrow f_{tmk}}^{(i-1)}\hat{x}_{x_{tn'k}\shortrightarrow f_{tmk}}^{(i-1)};}\\
{\scriptstyle \forall t,m,k,n:} & {\scriptstyle \tau_{f_{tmk}\shortrightarrow x_{tnk}}^{(i)}=\sigma_{\varpi}^{2}+\sum_{n'\neq n}\Bigl(\bigl|\hat{w}_{w_{mn'k}\shortrightarrow f_{tmk}}^{(i-1)}\bigr|^{2}\nu_{x_{tn'k}\shortrightarrow f_{tmk}}^{(i-1)}+\nu_{w_{mn'k}\shortrightarrow f_{tmk}}^{(i-1)}\bigl|\hat{x}_{x_{tn'k}\shortrightarrow f_{tmk}}^{(i-1)}\bigr|^{2}+\nu_{w_{mn'k}\shortrightarrow f_{tmk}}^{(i-1)}\nu_{x_{tn'k}\shortrightarrow f_{tmk}}^{(i-1)}\Bigr)}\\
{\scriptstyle \forall t,m,k,n:} & {\scriptstyle \hat{z}_{f_{tmk}\shortrightarrow w_{mnk}}^{(i)}=z_{f_{tmk}\shortrightarrow x_{tnk}}^{(i)}-\hat{x}_{x_{tnk}\shortrightarrow f_{tmk}}^{(i-1)}\hat{w}_{w_{mnk}}^{(i-1)};{\scriptstyle \hat{z}_{f_{tmk}\shortrightarrow x_{tnk}}^{(i)}=z_{f_{tmk}\shortrightarrow x_{tnk}}^{(i)}-\hat{w}_{w_{mnk}\shortrightarrow f_{tmk}}^{\left(i-1\right)}\hat{x}_{x_{tnk}}^{(i-1)};}}\\
{\scriptstyle \forall t,m,k,n:} & {\scriptstyle \hat{\tau}_{f_{tmk}\shortrightarrow w_{mnk}}^{(i)}=\tau_{f_{tmk}\shortrightarrow x_{tnk}}^{(i)}+\nu_{x_{tnk}\shortrightarrow f_{tmk}}^{(i-1)}\bigl|\hat{w}_{w_{mnk}}^{(i-1)}\bigr|^{2};}{\scriptstyle \hat{\tau}_{f_{tmk}\shortrightarrow x_{tnk}}^{(i)}=\tau_{f_{tmk}\shortrightarrow x_{tnk}}^{(i)}+\nu_{w_{mnk}\shortrightarrow f_{tmk}}^{(i-1)}\bigl|\hat{x}_{x_{tnk}}^{(i-1)}\bigr|^{2};}\\
{\scriptstyle \forall t,m,k,n:} & {\scriptstyle \nu_{f_{tmk}\shortrightarrow x_{tnk}}^{\left(i\right)}=\frac{\hat{\tau}_{f_{tmk}\shortrightarrow x_{tnk}}^{(i)}}{\bigl|\hat{w}_{w_{mnk}\shortrightarrow f_{tmk}}^{(i-1)}\bigr|^{2}+\nu_{w_{mnk}\shortrightarrow f_{tmk}}^{(i-1)}\biggl(1-\frac{\bigl|\hat{z}_{f_{tmk}\shortrightarrow x_{tnk}}^{(i)}\bigr|^{2}}{\hat{\tau}_{f_{tmk}\shortrightarrow x_{tnk}}^{(i)}}\biggr)};}{\scriptstyle \hat{x}_{f_{tmk}\shortrightarrow x_{tnk}}^{\left(i\right)}=\nu_{f_{tmk}\shortrightarrow x_{tnk}}^{(i)}\frac{\bigl(\hat{w}_{w_{mnk}\shortrightarrow f_{tmk}}^{(i-1)}\bigr)^{*}z_{f_{tmk}\shortrightarrow x_{tnk}}^{(i)}}{\hat{\tau}_{f_{tmk}\shortrightarrow x_{tnk}}^{(i)}};}\\
{\scriptstyle \forall t,m,k,n:} & {\scriptstyle \nu_{f_{tmk}\shortrightarrow w_{mnk}}^{\left(i\right)}=\frac{\hat{\tau}_{f_{tmk}\shortrightarrow w_{mnk}}^{(i)}}{\bigl|\hat{x}_{x_{tnk}\shortrightarrow f_{tmk}}^{(i-1)}\bigr|^{2}+\nu_{x_{tnk}\shortrightarrow f_{tmk}}^{(i-1)}\biggl(1-\frac{\hat{z}_{f_{tmk}\shortrightarrow w_{mnk}}^{(i)}\bigr|^{2}}{\hat{\tau}_{f_{tmk}\shortrightarrow w_{mnk}}^{(i)}}\biggr)};}{\scriptstyle \hat{w}_{f_{tmk}\shortrightarrow w_{mnk}}^{(i)}=\nu_{f_{tmk}\shortrightarrow w_{mnk}}^{(i)}\frac{\bigl(\hat{x}_{x_{tnk}\shortrightarrow f_{tmk}}^{(i-1)}\bigr)^{*}z_{f_{tmk}\shortrightarrow x_{tnk}}^{(i)}}{\hat{\tau}_{f_{tmk}\shortrightarrow w_{mnk}}^{(i)}};}\\
{\scriptstyle \forall t,n,k:} & {\scriptstyle {\scriptstyle \gamma_{x_{tnk}}^{(i)}}=\frac{1}{\sum_{m}\frac{1}{\nu_{f_{tmk}\shortrightarrow x_{tnk}}^{(i)}}};{\scriptstyle \zeta_{x_{tnk}}^{(i)}}=\gamma_{x_{tnk}}^{(i)}\sum_{m}\frac{\hat{x}_{f_{tmk}\shortrightarrow x_{tnk}}^{(i)}}{\nu_{f_{tmk}\shortrightarrow x_{tnk}}^{(i)}};\mu_{x_{tnk}\shortrightarrow\mathcal{M}_{tnk}}^{(i)}\left(x_{tnk}\right)=\mathcal{N}_{\mathbb{C}}\left(x_{tnk};\zeta_{x_{tnk}}^{(i)},\gamma_{x_{tnk}}^{(i)}\right);}\\
{\scriptstyle \forall t,n,k,q:} & {\scriptstyle \lambda_{e}^{(i)}(c_{tnk}^{q})=\mathsf{ln}\frac{\sum_{x_{tn}^{k}\in\mathcal{A}_{q}^{1}}\mu_{x_{tnk}\shortrightarrow\mathcal{M}_{tnk}}^{(i)}(x_{tnk})\mu_{\mathcal{M}_{tnk}\shortrightarrow x_{tnk}}^{(i-1)}(x_{tnk})}{\sum_{x_{tn}^{k}\in\mathcal{A}_{q}^{0}}\mu_{x_{tnk}\shortrightarrow\mathcal{M}_{tnk}}^{(i)}(x_{tnk})\mu_{\mathcal{M}_{tnk}\shortrightarrow x_{tnk}}^{(i-1)}(x_{tnk})}-\lambda_{a}^{(i-1)}(c_{tnk}^{q})}\\
{\scriptstyle \forall n:} & {\scriptstyle \mathsf{Decode}\thinspace\mathsf{and}\thinspace\mathsf{generate}\thinspace\mathsf{LLRs\thinspace}\left\{ \lambda_{a}^{(i)}(c_{tnk}^{q}),\forall t,\forall k,\forall q\right\} }\\
{\scriptstyle \forall t,n,k:} & {\scriptstyle \mu_{\mathcal{M}_{tnk}\shortrightarrow x_{tnk}}^{(i)}(x_{tnk})=\prod_{q}\frac{\mathsf{exp}\bigl(c_{tnk}^{q}\lambda_{a}^{(i)}(c_{tnk}^{q})\bigr)}{1+\mathsf{exp}\bigl(c_{tnk}^{q}\lambda_{a}^{(i)}(c_{tnk}^{q})\bigr)};{\scriptstyle {\scriptstyle \beta_{x_{tnk}}^{(i)}\left(x_{tnk}\right)=\frac{\mu_{\mathcal{M}_{tnk}\shortrightarrow x_{tnk}}^{(i)}\left(x_{tnk}\right)\mu_{x_{tnk}\shortrightarrow\mathcal{M}_{tnk}}^{(i)}\left(x_{tnk}\right)}{\sum_{x_{tnk}\in\mathcal{A}}\mu_{\mathcal{M}_{tnk}\shortrightarrow x_{tnk}}^{(i)}(x_{tnk})\mu_{x_{tnk}\shortrightarrow\mathcal{M}_{tnk}}^{(i)}(x_{tnk})}}};}\\
{\scriptstyle \forall t,n,k:} & {\scriptstyle {\scriptstyle \hat{x}_{x_{tnk}}^{(i)}}=\sum_{\alpha_{s}\in\mathcal{A}}\alpha_{s}\beta_{x_{tnk}}^{(i)}\left(x_{tnk}=\alpha_{s}\right);}{\scriptstyle \nu_{x_{tnk}}^{(i)}}{\scriptstyle =\sum_{\alpha_{s}\in\mathcal{A}}\left|\alpha_{s}\right|^{2}\beta_{x_{tnk}}^{(i)}\left(x_{tnk}=\alpha_{s}\right)-\bigl|\hat{x}_{tnk}^{(i)}\bigr|^{2};}\\
{\scriptstyle \forall t,n,k,m:} & {\scriptstyle \nu_{x_{tnk}\shortrightarrow f_{tmk}}^{(i)}}{\scriptstyle =\nu_{x_{tnk}}^{(i)}\frac{\nu_{f_{tmk}\shortrightarrow x_{tnk}}^{(i)}}{\nu_{f_{tmk}\shortrightarrow x_{tnk}}^{(i)}-\nu_{x_{tnk}}^{(i)}},\forall m;}{\scriptstyle \hat{x}_{x_{tnk}\shortrightarrow f_{tmk}}^{(i)}=\hat{x}_{x_{tnk}}^{\left(i\right)}+\nu_{x_{tnk}}^{(i)}\frac{\hat{x}_{x_{tnk}}^{\left(i\right)}-\hat{x}_{f_{tmk}\shortrightarrow x_{tnk}}^{(i)}}{\nu_{f_{tmk}\shortrightarrow x_{tnk}}^{(i)}-\nu_{x_{tnk}}^{(i)}}.}
\end{align*}
\rule{1\linewidth}{0.5pt} \caption{The EP-QA at the $i\text{th}$ turbo iteration.}
\label{alg:EP-QA}
\end{table}

Summing up the above discussions, the EP based message passing for
the\emph{ detection-decoding-loop} is formulated in Table \ref{alg:EP-QA},
which will be referred to as ``EP-QA''. At the first turbo iteration,
we set $\hat{x}_{x_{tnk}\shortrightarrow f_{tmk}}^{(0)}=0,\nu_{x_{tnk}\shortrightarrow f_{tmk}}^{(0)}=1,\forall t,n,k,m$;
$\hat{w}_{w_{mnk}\shortrightarrow f_{tmk}}^{(0)}=0,\nu_{w_{mnk}\shortrightarrow f_{tmk}}^{(0)}=1,\forall t,m,k,n$;
and $\lambda_{a}^{(0)}(c_{tnk}^{q})=0,\forall t,n,k,q$. When updating
messages, it can be observed that the variance parameters $\nu_{f_{tmk}\shortrightarrow w_{mnk}}^{(i)}$
in (\ref{eq:nu_=00007Bf_=00007Btmk=00007D2w_=00007Bmnk=00007D=00007D^=00007B(i)=00007D})
and $\nu_{x_{tnk}\shortrightarrow f_{tmk}}^{(i)}$ in (\ref{eq:nu_=00007Bx_=00007Btnk=00007D2f_=00007Btmk=00007D=00007D^=00007B(i)=00007D})
take negative values in rare situations, which can lead to erratic
behavior and is common in many EP implementations \cite{minka2001family}.
To circumvent this problem, we change a negative $\nu_{f_{tmk}\shortrightarrow w_{mnk}}^{(i)}$
to $+\infty$ ( a large positive constant is used in practice, e.g.,
$10^{6}$, see \cite{andersen2015bayesian} and \cite{hernandez2013generalized})
and change a negative $\nu_{x_{tnk}\shortrightarrow f_{tmk}}^{(i)}$
to $\nu_{x_{tnk}}^{(i)}$ shown in (\ref{eq:nu_=00007Bx_=00007Btnk=00007D=00007D^=00007B(i)=00007D}).
Although this is just a heuristic measure, in our simulations it indeed
avoids the instability of expectation propagation.

\subsection{Message Updating in Channel-Estimation-Loop}

\begin{table}
\rule{1\linewidth}{0.5pt}

$\begin{aligned}{\scriptstyle \forall m,n,k:} & {\scriptstyle z_{g_{mnk}}^{(i)}=\hat{w}_{w_{mnk}\shortrightarrow g_{mnk}}^{(i)}-\sum_{l}\phi_{kl}\hat{h}_{h_{mnl}}^{(i-1)}+\epsilon_{mnk}^{(i-1)};\tau_{g_{mnk}}^{(i)}=\nu_{w_{mnk}\shortrightarrow g_{mnk}}^{(i)}+\sum_{l}\nu_{h_{mnl}}^{(i-1)}.}\\
{\scriptstyle \forall mn:} & {\scriptstyle \bar{\tau}_{mn}^{(i)}=\sum_{k}\tau_{g_{mnk}}^{(i)}/K.}\\
{\scriptstyle \forall m,n,l:} & {\scriptstyle \xi_{mnl}^{(i)}=\sum_{k}\frac{\varPhi_{kl}^{*}z_{g_{mnk}}^{(i)}}{\tau_{g_{mnk}}^{(i)}}+\hat{h}_{h_{mnl}}^{(i-1)}\sum_{k'}\frac{1}{\tau_{g_{mnk'}}^{(i)}}-\frac{\nu_{h_{mnl}}^{(i-1)}}{\bar{\tau}_{mn}^{(i)}}\xi_{mnl}^{(i-1)};\nu_{h_{mnl}}^{(i)}=\frac{1}{\frac{M}{\sum_{m'}\biggl(\Bigl|\hat{h}_{h_{m'nl}}^{(i-1)}\Bigr|^{2}+\nu_{h_{m'nl}}^{(i-1)}\biggr)}+\sum_{k}\frac{1}{\tau_{g_{mnk}}^{(i)}}};\hat{h}_{h_{mnl}}^{(i)}=\nu_{h_{mnl}}^{(i)}\xi_{mnl}^{(i)}.}\\
{\scriptstyle \forall mn:} & {\scriptstyle \bar{\nu}_{mn}^{(i)}=\sum_{l}\nu_{h_{mnl}}^{(i)}/L.}\\
{\scriptstyle \forall m,n,k:} & {\scriptstyle \epsilon_{mnk}^{(i)}=\frac{z_{g_{mnk}}^{(i)}\sum_{l}\nu_{h_{mnl}}^{(i)}+\sum_{l'}\phi_{kl'}\nu_{h_{mnl'}}^{(i)}\hat{h}_{h_{mnl'}}^{(i-1)}-\bar{\nu}_{mn}^{(i)}\epsilon_{mnk}^{(i-1)}}{\tau_{g_{mnk}}^{(i)}};{\scriptstyle \hat{w}_{g_{mnk}\shortrightarrow w_{mnk}}^{(i)}=\sum_{l}\phi_{kl}\hat{h}_{h_{mnl}}^{(i)}-\epsilon_{mnk}^{(i)};}\nu_{g_{mnk}\shortrightarrow w_{mnk}}^{(i)}=\sum_{l}\nu_{h_{mnl}}^{(i)}.}\\
{\scriptstyle \forall m,n,k:} & {\scriptstyle \nu_{w_{mnk}}^{(i)}=\frac{1}{\frac{1}{\nu_{g_{mnk}\shortrightarrow w_{mnk}}^{(i)}}+\sum_{t}\frac{1}{\nu_{f_{tmk}\shortrightarrow w_{mnk}}^{(i)}}},\hat{w}_{w_{mnk}}^{(i)}=\nu_{w_{mnk}}^{(i)}\Biggl(\frac{\hat{w}_{g_{mnk}\shortrightarrow w_{mnk}}^{(i)}}{\nu_{g_{mnk}\shortrightarrow w_{mnk}}^{(i)}}+\sum_{t}\frac{\hat{w}_{f_{tmk}\shortrightarrow w_{mnk}}^{(i)}}{\nu_{f_{tmk}\shortrightarrow w_{mnk}}^{(i)}}\Biggr).}\\
{\scriptstyle \forall m,n,k,t:} & {\scriptstyle \nu_{w_{mnk}\shortrightarrow f_{tmk}}^{(i)}=\frac{1}{\frac{1}{\nu_{w_{mnk}}^{(i)}}-\frac{1}{\nu_{f_{tmk}\shortrightarrow w_{mnk}}^{(i)}}},\hat{w}_{w_{mnk}\shortrightarrow f_{tmk}}^{(i)}=\nu_{w_{mnk}\shortrightarrow f_{tmk}}^{(i)}\Biggl(\frac{\nu_{w_{mnk}}^{(i)}}{\hat{w}_{w_{mnk}}^{(i)}}-\frac{\hat{w}_{f_{tmk}\shortrightarrow w_{mnk}}^{(i)}}{\nu_{f_{tmk}\shortrightarrow w_{mnk}}^{(i)}}\Biggr).}
\end{aligned}
$

\rule{1\linewidth}{0.5pt}\caption{The GMP at the $i\text{th}$ turbo iteration.}

\label{alg:GMP}
\end{table}

Now we focus on Bayesian learning of the hyper-parameters $\{\gamma_{nl}\}$,
as it is unknown to the receiver. Using the variational message-passing
rule \cite{winn2005variational}, we obtain the message from the function
node $\psi_{mnl}$ to the precision variable $\gamma_{nl}$,
\begin{equation}
\mu_{\psi_{mnl}\shortrightarrow\gamma_{nl}}^{(i)}(\gamma_{nl})=\mathsf{exp}\biggl(\mathsf{E}_{\beta_{h_{mnl}}^{(i-1)}\left(h_{mnl}\right)}\mathsf{ln}\psi_{mnl}\left(h_{mnl},\gamma_{nl}\right)\biggr)\propto\mathsf{Gam}\Bigl(\gamma_{nl};0,\bigl|\hat{h}_{h_{mnl}}^{(i-1)}\bigr|^{2}+\nu_{h_{mnl}}^{(i-1)}\Bigr),
\end{equation}
where $\beta_{h_{mnl}}^{(i-1)}(h_{mnl})=\mathcal{N}_{\mathbb{C}}\Bigl(h_{mnl};\hat{h}_{h_{mnl}}^{(i-1)},\nu_{h_{mnl}}^{(i-1)}\Bigr)$
is the belief of $h_{mnl}$ at the $\left(i-1\right)\text{th}$ turbo
iteration. Then the belief of precision variable $\gamma_{nl}$ is
updated by
\begin{equation}
\beta_{\gamma_{nl}}^{(i)}(\gamma_{nl})=p(\gamma_{nl})\prod_{m}\mu_{\psi_{mnl}\shortrightarrow\gamma_{nl}}^{(i)}(\gamma_{nl})\propto\mathsf{Gam}\biggl(\gamma_{nl};M,\sum_{m}\Bigl(\bigl|\hat{h}_{h_{mnl}}^{(i-1)}\bigr|^{2}+\nu_{h_{mnl}}^{(i-1)}\Bigr)\biggr),
\end{equation}
Using the variational message-passing rule again, the message from
the function node $\psi_{mnl}(h_{mnl},\gamma_{nl})$ to variable node
$h_{mnl}$ reads
\begin{equation}
\mu_{\psi_{mnl}\shortrightarrow h_{mnl}}^{(i)}(h_{mnl})=\mathsf{exp}\left(\mathsf{E}_{\beta_{\gamma_{nl}}^{(i)}(\gamma_{nl})}\mathsf{ln}\psi_{mnl}(h_{mnl},\gamma_{nl})\right)\propto\mathcal{N}_{\mathbb{C}}\biggl(h_{mnl};0,\frac{1}{M}\sum_{m}\Bigl(\bigl|\hat{h}_{h_{mnl}}^{(i-1)}\bigr|^{2}+\nu_{h_{mnl}}^{(i-1)}\Bigr)\biggr),\label{eq:mu_=00007Bpsi_=00007Bmnl=00007D2h_=00007Bmnl=00007D=00007D^=00007B(i)=00007D}
\end{equation}
and the belief of $h_{mnl}$ is updated by $\beta_{h_{mnl}}^{(i)}\left(h_{mnl}\right)=\mu_{\psi_{mnl}\shortrightarrow h_{mnl}}^{(i)}(h_{mnl})\prod_{k}\mu_{g_{mnk}\shortrightarrow h_{mnl}}^{(i)}(h_{mnl})$,
where $\mu_{g_{mnk}\shortrightarrow h_{mnl}}^{(i)}(h_{mnl})$ is the
message from $g_{mnk}$ to $h_{mnl}$.

Following the derivation in \cite{Wu1405:Expectation}, the Gaussian
message passing for channel-estimation task, i.e., updating $\{\nu_{w_{mnk}\shortrightarrow f_{tmk}}^{(i)},\hat{w}{}_{w_{mnk}\shortrightarrow f_{tmk}}^{(i)}\}$,
is given by Table \ref{alg:GMP}, which will be referred to as ``GMP''.
At the first turbo iteration, i.e., $i=1$, we set $\bigl|\hat{h}_{h_{mnl}}^{(0)}\bigr|^{2}+\nu_{h_{mnl}}^{(0)}=1/L,\forall m,n,l$,
and $\xi_{mnl}^{(0)}=0,\forall m,n,l$.

\section{Complexity Comparisons\label{sec:Complexity-Comparisons}}

\begin{table}[h]
\caption{Complexity of detection and decoding per turbo iteration in terms
of FLOPs.}

\centering%
\begin{tabular}{cc}
\hline
Algorithm & FLOPs per Iteration\tabularnewline
\hline
EP-QA-L / EP-QA & $47TMNK+(11N+4)M\left(K-K_{p}\right)+\left(23\left|\mathcal{A}\right|+3Q\left|\mathcal{A}\right|+Q\right)TNK$\tabularnewline
BP-GA \cite{liu2009joint,novak2013idma} & $\left(28\left|\mathcal{A}\right|+33\right)TMNK+(2\left|\mathcal{A}\right|+3Q\left|\mathcal{A}\right|+Q)TNK$\tabularnewline
BP-MF\cite{Carles2011} & $19TMNK+(11N+4)M\left(K-K_{p}\right)+(23\left|\mathcal{A}\right|+3Q\left|\mathcal{A}\right|+Q)TNK$\tabularnewline
BP-MF-M \cite{Badiu2013} & $33TMNK+(11N+4)M\left(K-K_{p}\right)+(23\left|\mathcal{A}\right|+3Q\left|\mathcal{A}\right|+Q)TNK$\tabularnewline
\hline
\end{tabular}\label{table:d=000026d_complexity}
\end{table}

\begin{table}[h]
\caption{Complexity of channel estimation per turbo iteration in terms of FLOPs.}

\centering%
\begin{tabular}{cc}
\hline
Algorithm & FLOPs per Iteration\tabularnewline
\hline
EP-QA-L & $MN(20K\mathsf{log}_{2}K+30TK+11K-26TK_{p}+13K_{p}+18L-2)$\tabularnewline
EP-QA / BP-GA & $MN(20K\mathsf{log}_{2}K+30TK+11K-26TK_{p}+13K_{p}+14L-2)$\tabularnewline
BP-MF \cite{Carles2011} & $MN(16K^{3}+12K^{2}+17TK-K)+2TNK-2NK-2MN$\tabularnewline
BP-MF-M \cite{Badiu2013} & $MN(118G^{2}+68G-4)K-112G^{3}-92G^{3}+5G$\tabularnewline
\hline
\end{tabular}\label{table:channel_estimation_complexity}
\end{table}
\begin{figure}
\centering\includegraphics[width=3.5in]{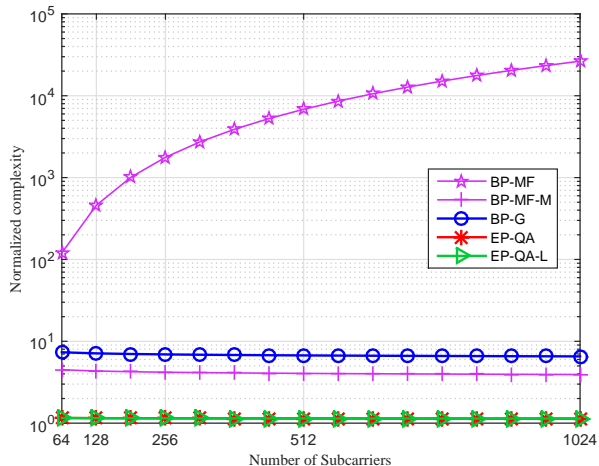}\caption{Normalized complexity of joint algorithms versus the number of subcarriers
$K$ in a $64\times8$ MIMO-OFDM systems with 16QAM, where $K_{p}=L=K/8$,
and $T=8$. The complexity is normalized over the complexity of joint
algorithm EP-QA-L. }

\label{fig:complexity_QPSK}
\end{figure}
In the following, EP-QA-L denotes the joint algorithm using the EP-QA
shown in Table \ref{alg:EP-QA} to complete detection and decoding
and using the GMP shown in Table \ref{alg:GMP} to complete channel
estimation; EP-QA and BP-GA denote the joint algorithms using the
EP-QA shown in Table \ref{alg:EP-QA} and the BP-GA ((\ref{eq:y_tm^k})-(\ref{eq:eq:m_fjk2gjik-1-1}))
to complete detection and decoding, respectively, and both using the
GMP with oracle PDP to complete channel estimation, i.e., the term
$\frac{1}{M}\sum_{m}\bigl(\bigl|\hat{h}_{h_{mnl}}^{(i-1)}\bigr|^{2}+\nu_{h_{mnl}}^{(i-1)}\bigr)$
in (\ref{eq:mu_=00007Bpsi_=00007Bmnl=00007D2h_=00007Bmnl=00007D=00007D^=00007B(i)=00007D})
is replaced by the true path power of $h_{mnl}$; the BP-MF denotes
the BP-MF algorithm employing disjoint channel model proposed in \cite{Carles2011}
and \cite{Riegler2013}; and BP-MF-M denotes the low-complexity version
of BP-MF algorithm employing Markov channel model proposed in \cite{Badiu2013}.
Note that, both the BP-MF and the BP-MF-M require prior knowledge
of the channel PDP. We compare the complexity of our proposed EP-QA-L
algorithm with that of the EP-QA, the BP-GA, the BP-MF, and the BP-MF-M.
The complexity is evaluated in terms of floating-point operations
(FLOPs) per iteration. Here we do not distinguish the complexity of
addition, subtraction, multiplication, and division for simplicity.
Note that the multiplication of a complex number and a real number
needs two FLOPs, and the multiplication of two complex numbers (excluding
conjugate numbers) needs six FLOPs. It is assumed that the operation
of $\mathsf{exp}\left(\cdot\right)$ can be implemented by a look-up
table and $\{\lambda_{e}^{(i)}(c_{tnk}^{q})\}$ is calculated by the
decoders, which are not taken into account. Table \ref{table:d=000026d_complexity}
shows the complexity of these algorithms performing detection and
decoding. For channel estimation, the complexity is listed in Table
\ref{table:channel_estimation_complexity}. The normalized complexity
of these joint algorithms per turbo iteration versus number of subcarriers
$K$ in a $64\times8$ MIMO-OFDM systems with 16QAM is shown in Fig
\ref{fig:complexity_QPSK}, where $K_{p}=L=K/8$ and $T=8$. The EP-QA-L
and EP-QA have almost the same complexity, while the EP-Q has the
lowest complexity. As the number of subcarriers increases from 64
to 1024, the complexity of EP-QA-L is about $\frac{1}{3}$ of that
of BP-MF-M, about $\frac{1}{6}$ of that of BP-G, and about $\frac{1}{100}\sim\frac{1}{23000}$
of that of BP-MF.

\section{Simulation Results \label{sec:Simulation-Results}}

The proposed EP-QA-L is compared with the EP-QA, the BP-MF, the BP-MF-M,
and the BP-GA in terms of normalized mean square error (NMSE) of the
channel weights and BER, as well as the matched filter bound (MFB)
that is obtained by the MAP decoding under the condition of perfect
multiuser interference cancellation and perfect channel state information
(PCSI).

Due to space constraints, a selected set of system parameters is used
for simulations\footnote{We will make our simulation package available for download after (possible)
acceptance of the paper.}. We consider the uplink of a multiuser system with $N=8$ independent
users, and each user is equipped with one transmit antenna. For each
user, the transmission is based on OFDM with $K=128$ subcarriers.
We choose a $R=1/2$ recursive systematic convolutional (RSC) code
with generator polynomial $\left[G_{1},G_{2}\right]=\left[117,155\right]_{\text{\ensuremath{\mathsf{oct}}}}$,
followed by a random interleaver. For bit-to-symbol mapping, multilevel
Gray-mapping is used \cite{schniter2011message}. The maximum multipath
delay $L=16$ is assumed and the PDP is modeled as exponentially decaying,
i.e., $\gamma_{nl}=\frac{e^{-l/6}}{\sum_{l=1}^{L}e^{-l/6}},\forall n$.
The CP length is set to be $L_{\mathsf{cp}}=L$ and the pilot length
is also set to be $K_{p}=16$. We adopt the channel model in (\ref{eq:kronecker_channel})
with the spatial correlation matrix in (\ref{eq:kronecker_approx}).
Considering a massive ($16\times4$) UPA and a moderate ($16\times1)$
unformed linear array (ULA), we set the antenna spacing to $d_{\mathsf{az}}=d_{\mathsf{el}}=\lambda$,
uniformly generate following random variables independently for each
user in a channel realization: the mean of horizontal AoD $\theta_{\mathsf{az}}$
in $[\pi/6,5\pi/6)$, the mean of vertical AoD $\theta_{\mathsf{el}}$
in $[\pi/12,\pi/3)$, and the standard deviations of horizontal AoD
$\sqrt{\nu_{\mathsf{az}}}$ and vertical AoD $\sqrt{\nu_{\mathsf{el}}}$
both in $[\pi/12,\pi/6)$. At the receiver, the BCJR algorithm is
used to decode the convolutional codes. The channels are assumed to
be block-static for the selected $T=8$ transmitted OFDM symbols.

Taking into account of the overhead incurred by the CP and the frequency-domain
pilots, the spectral efficiency $\eta$ of the MIMO-OFDM scheme normalized
by the ideal case without any overhead is expressed as $\eta=\frac{TNK-N^{2}K_{p}}{TN(L_{\mathsf{cp}}+K)}=77.8\%$
\cite{Dai2013Spectrally}. The energy per bit to noise power spectral
density ratio $E_{b}/N_{0}$ is defined as \cite{Hochwald2003}
\begin{equation}
\frac{E_{b}}{N_{0}}=\frac{E_{s}}{N_{0}}+10\log_{10}\frac{M}{\eta RNQ},
\end{equation}
where $E_{s}/N$ is the average energy per transmitted symbol. For
a fixed $E_{b}/N_{0}$, then $E_{s}/N_{0}$ is scaled down by the
number of receive antennas $M$.

\begin{figure}
\centering\includegraphics[width=3.5in]{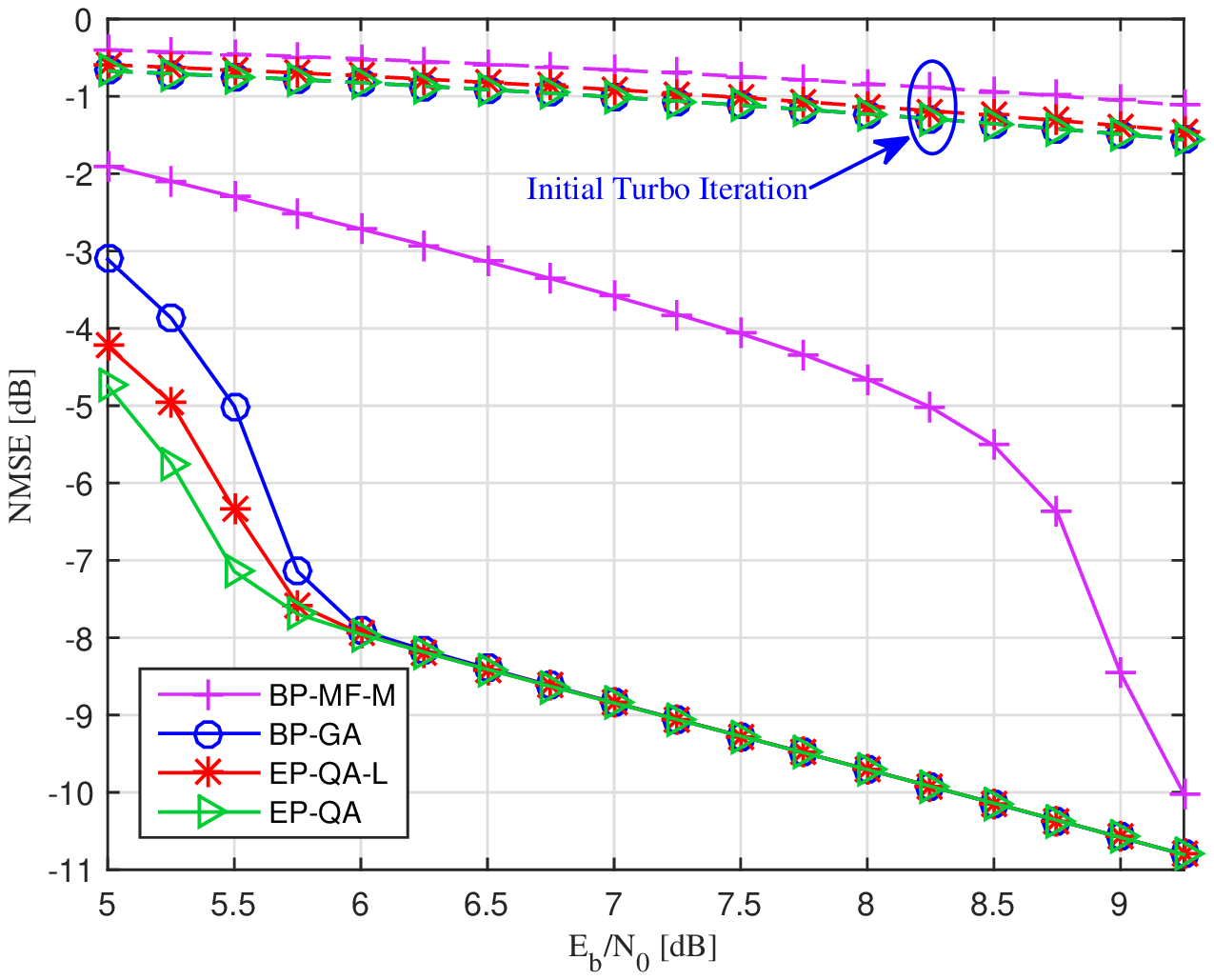}\caption{NMSE versus $E_{b}/N_{0}$ in the $64\times8$ MIMO system with 16QAM. }

\label{fig:MSE_64_8_16QAM}
\end{figure}

\begin{figure}
\centering\includegraphics[width=3.5in]{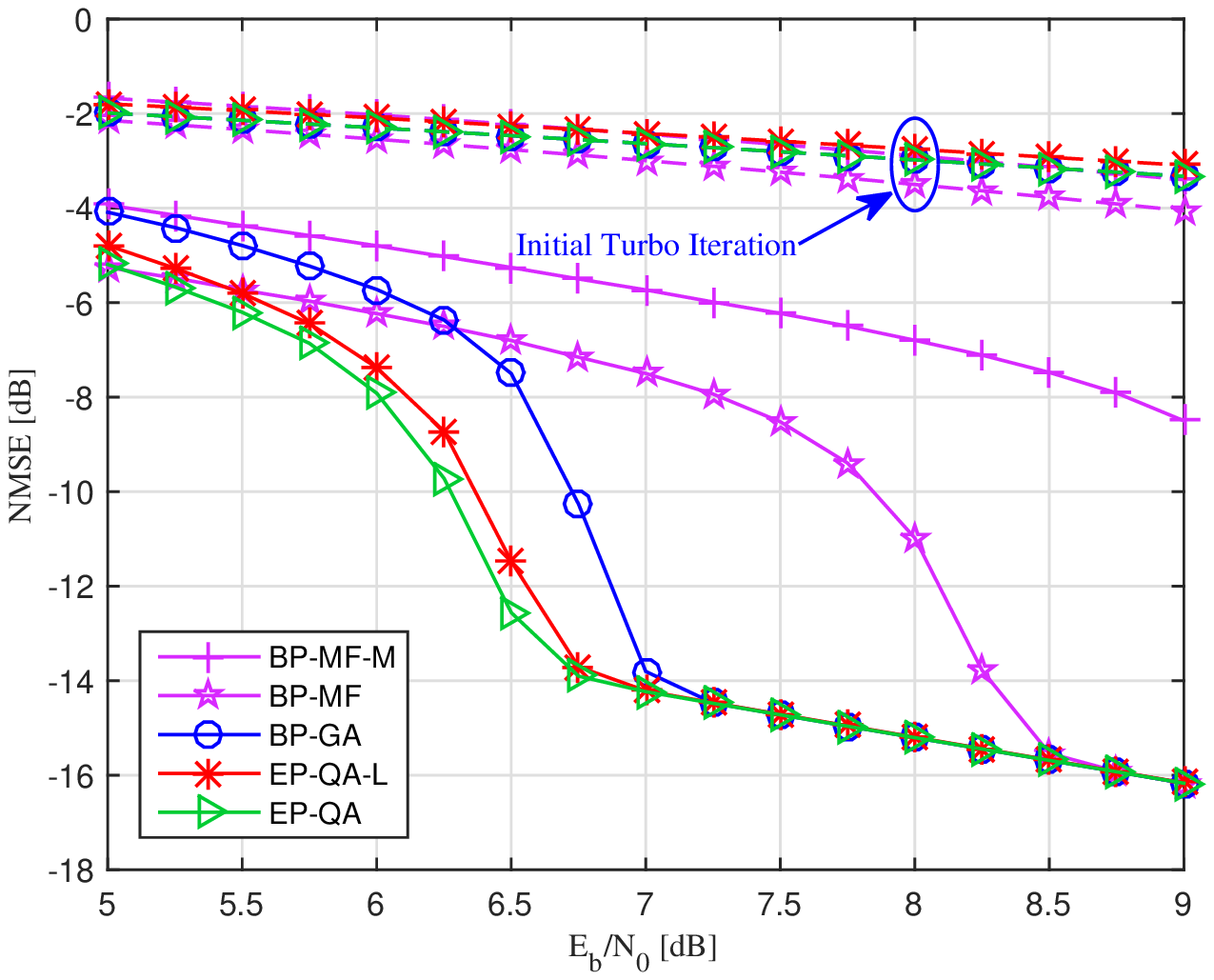}\caption{NMSE versus $E_{b}/N_{0}$ in the $16\times8$ MIMO system with 16QAM. }

\label{fig:MSE_16_8_16QAM}
\end{figure}

\begin{figure}
\centering\subfloat[BP-MF-M.]{\centering\includegraphics[width=3.25in]{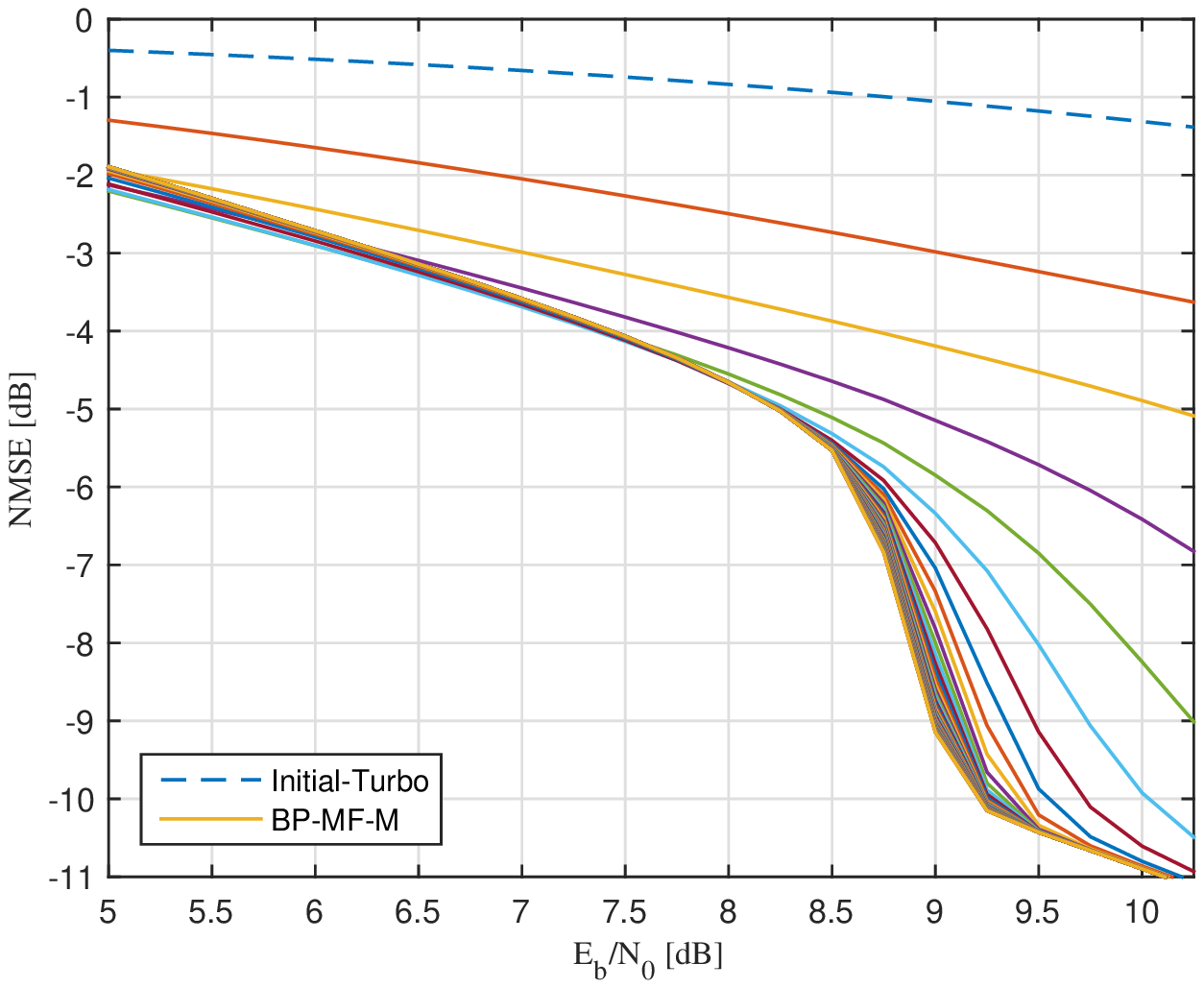}}\subfloat[BP-GA.]{\centering\includegraphics[width=3.25in]{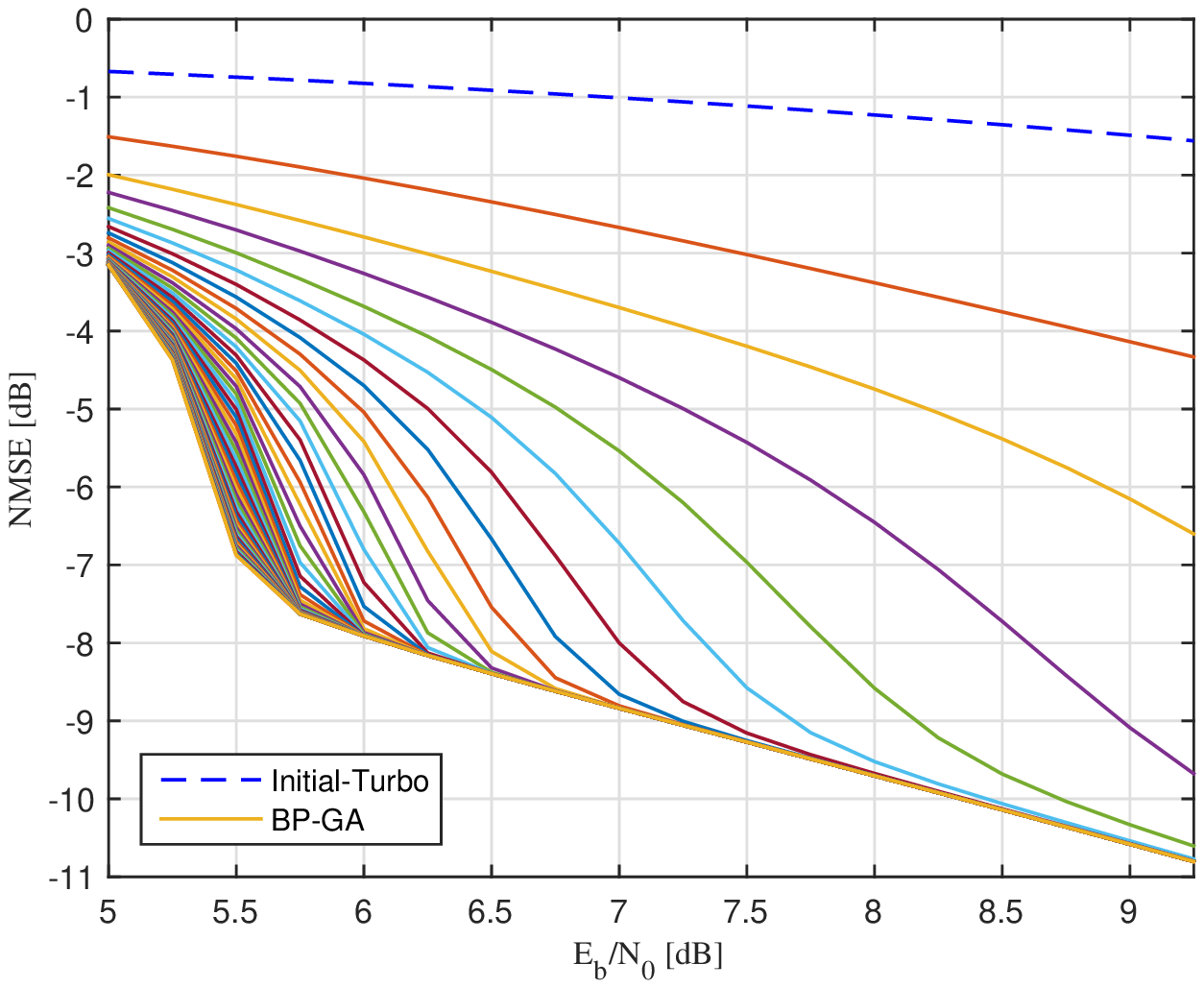}}

\subfloat[EP-QA-L.]{\centering\includegraphics[width=3.25in]{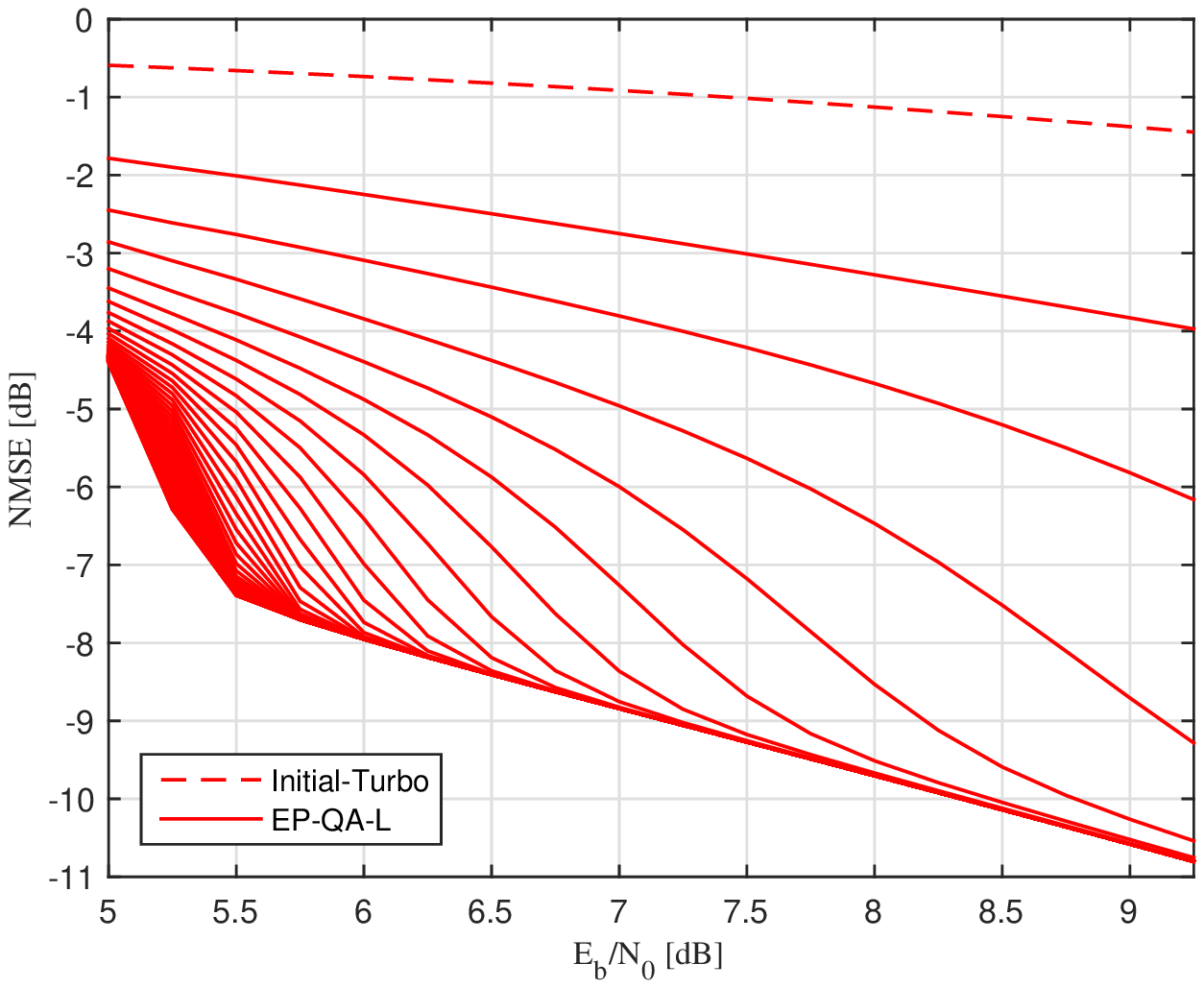}}\subfloat[EP-QA.]{\centering\includegraphics[width=3.25in]{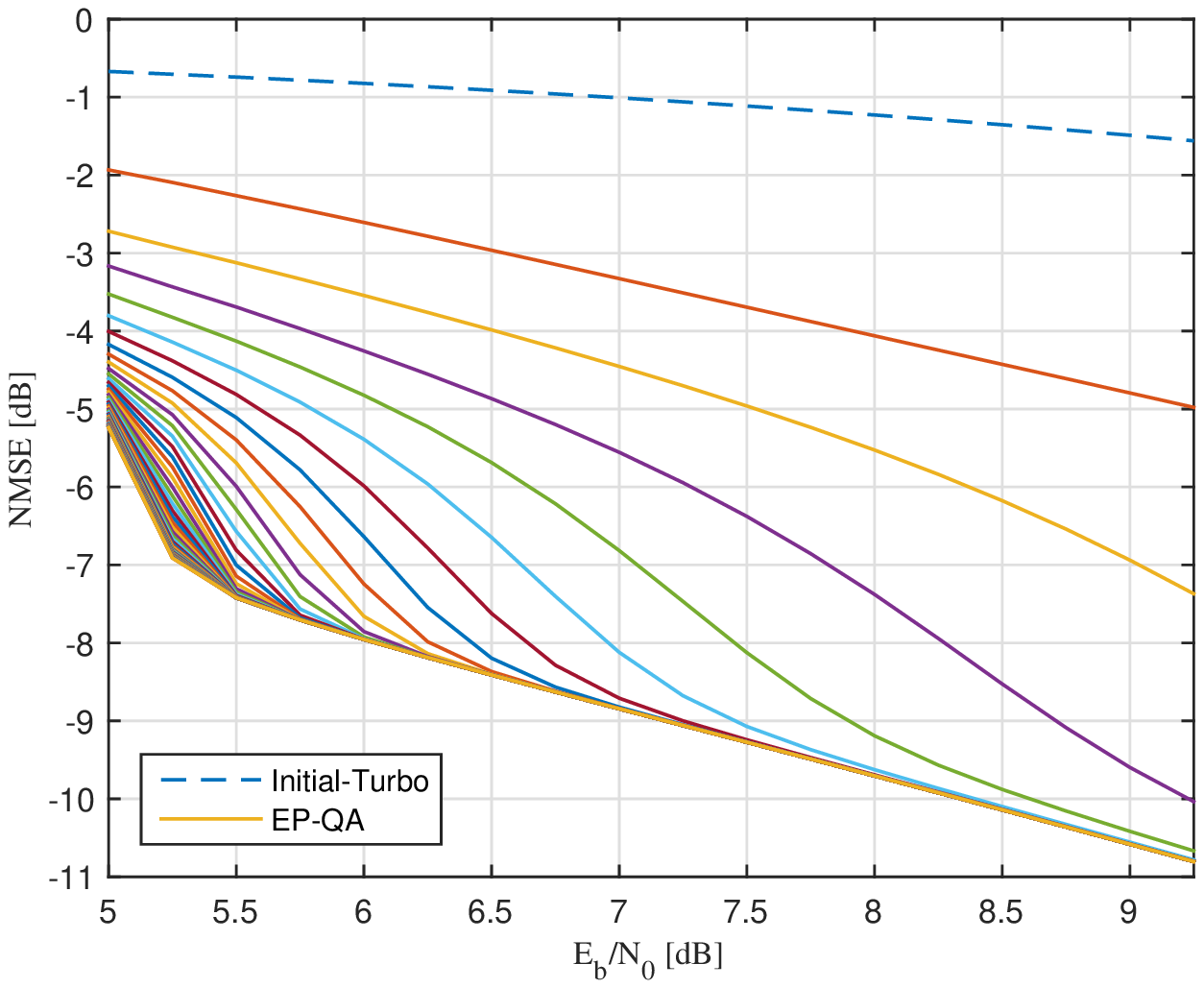}}

\caption{NMSE versus $E_{b}/N_{0}$ with multiple iterations in the $64\times8$
MIMO system with 16QAM. }
\label{fig:MSE_64_8_64QAM_Itera}
\end{figure}

\begin{figure}
\centering\subfloat[BP-MF-M.]{\centering\includegraphics[width=3.25in]{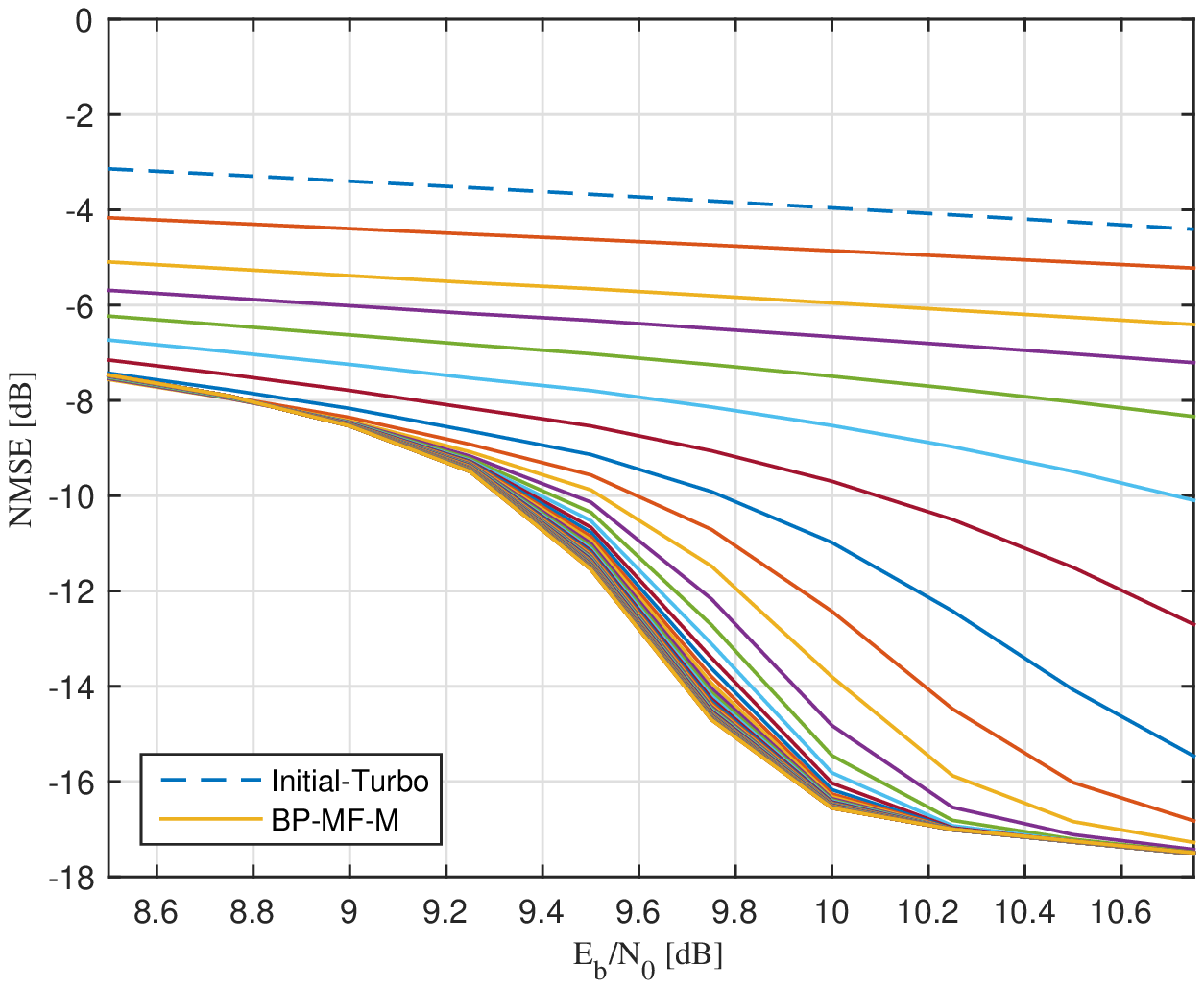}}\subfloat[BP-MF.]{\centering\includegraphics[width=3.25in]{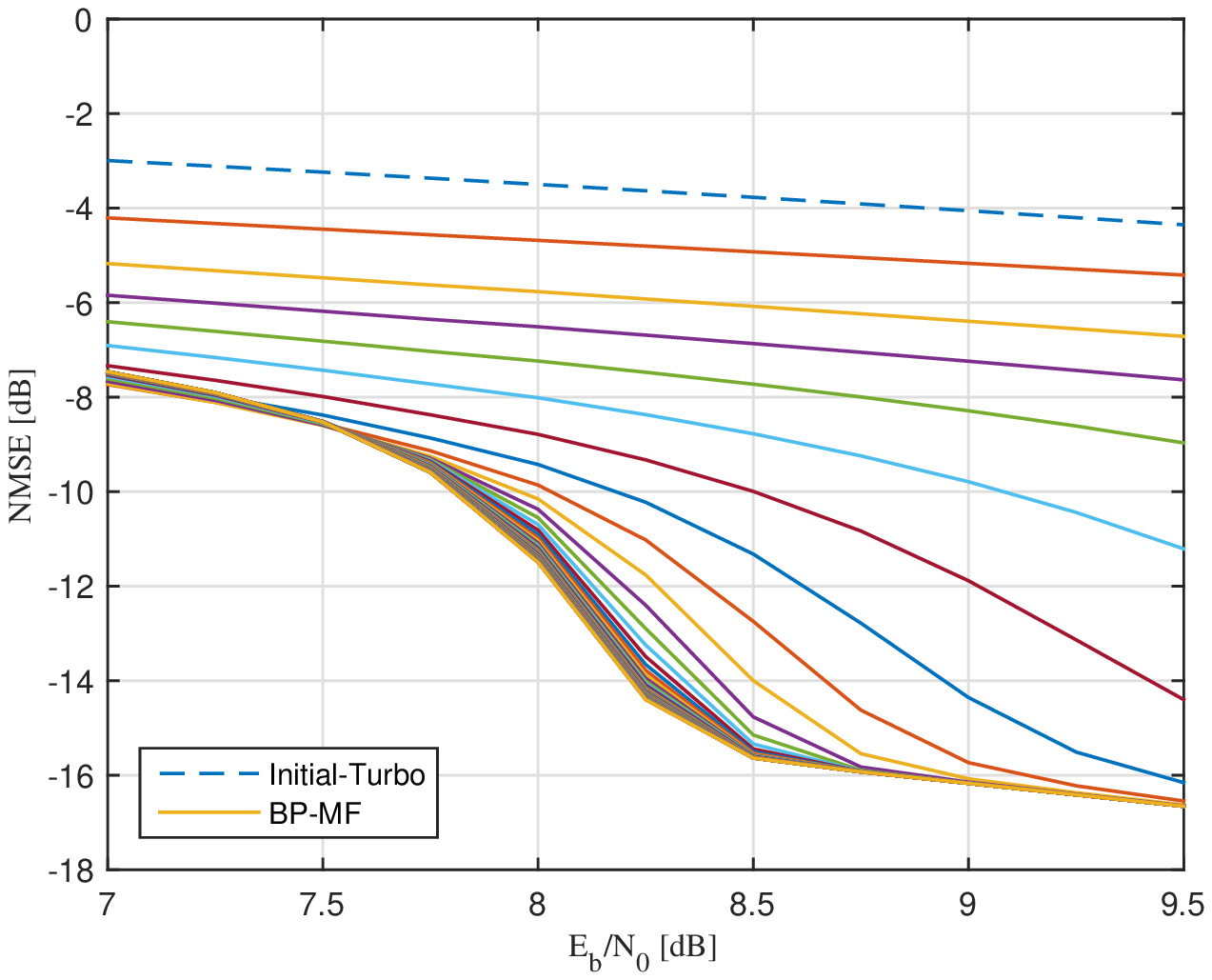}}

\centering\subfloat[BP-GA.]{\centering\includegraphics[width=3.25in]{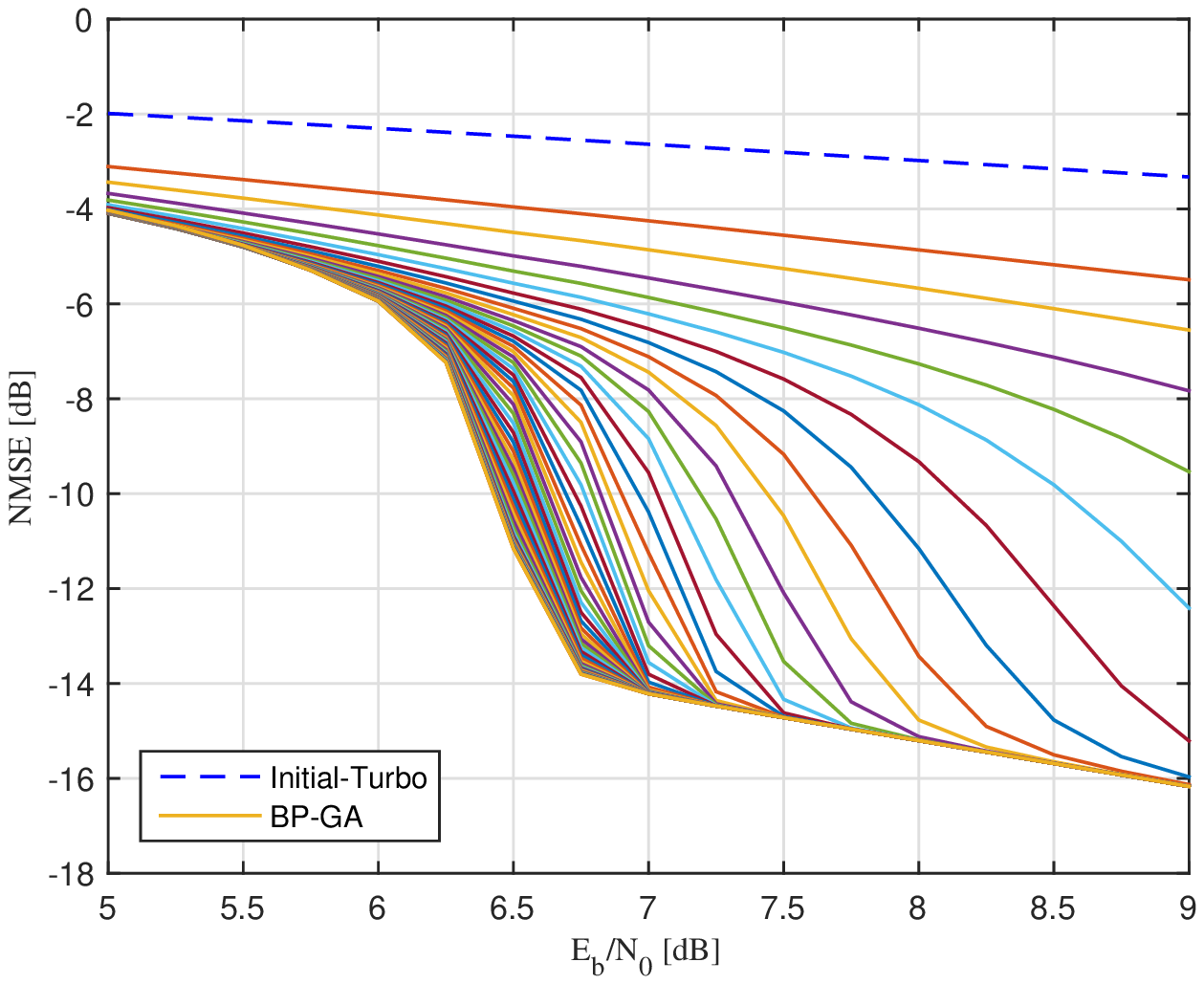}}\subfloat[EP-QA-L.]{\centering\includegraphics[width=3.25in]{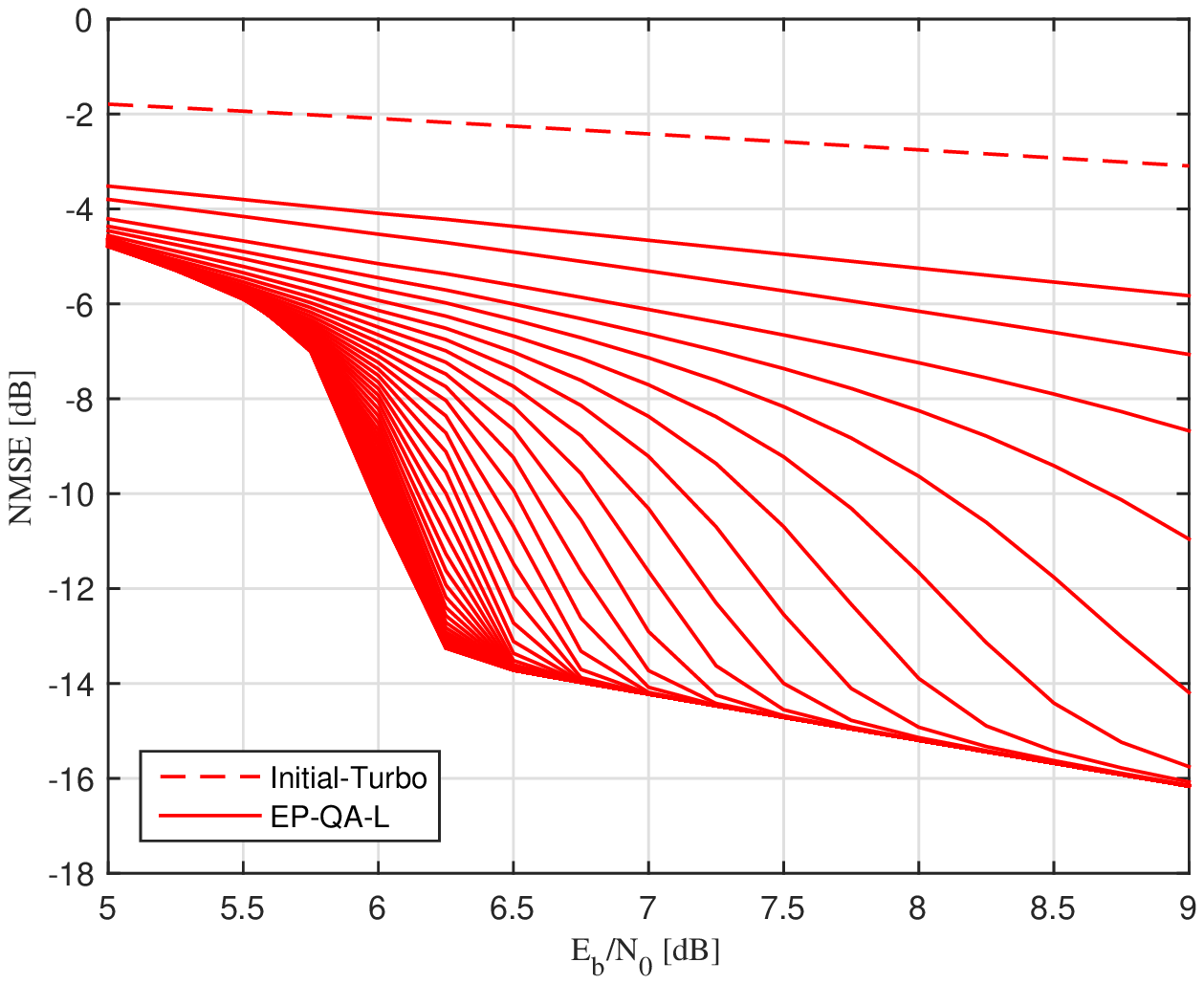}}

\centering\subfloat[EP-QA.]{\centering\includegraphics[width=3.25in]{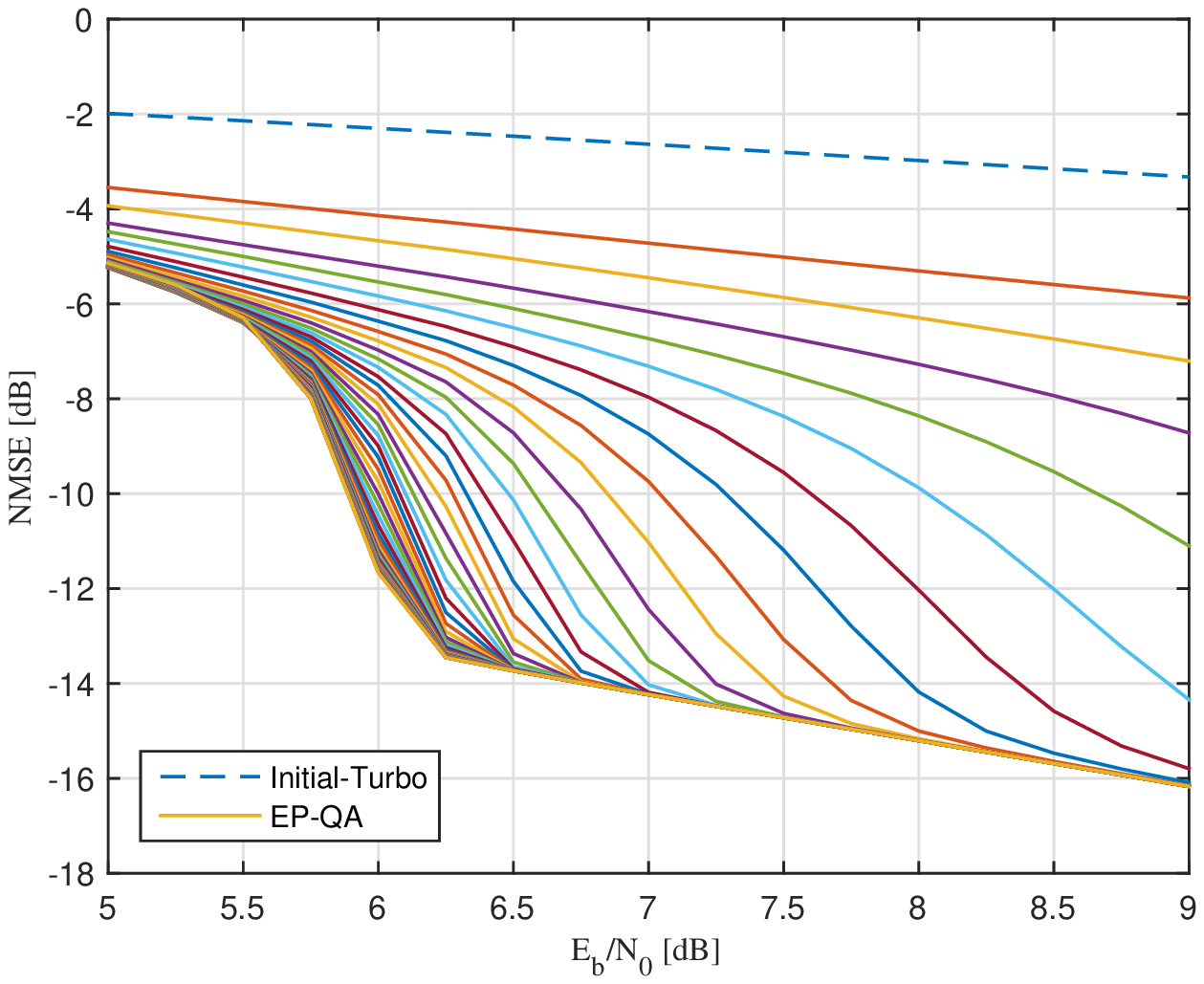}}

\caption{NMSE versus $E_{b}/N_{0}$ with multiple iterations in the $16\times8$
MIMO system with 16QAM. }
\label{fig:MSE_16_8_16QAM_Itera}
\end{figure}

\subsection{Channel-Tap $NMSE$ Versus $E_{b}/N_{0}$}

In the initial turbo iteration, only the pilot symbols are available
for the channel estimation. For the EP-QA-L and the EP-QA, the channel
estimation loops perform 5 inner iterations in the initial turbo iteration
and perform only 1 inner iteration in each of the following turbo
iterations. For the BP-MF, the channel estimator is equivalent to
a pilot-based LMMSE estimator in the initial turbo iteration, and
becomes a data-aided LMMSE in the next turbo iterations. The channel
estimation of the BP-MF-M is performed by a Kalman smoother proposed
in \cite{Badiu2013}, where the group-size of contiguous channel weights
is set to be $G=4$.

Fig. \ref{fig:MSE_64_8_16QAM} and Fig. \ref{fig:MSE_16_8_16QAM}
show the NMSE of the channel estimation versus $E_{b}/N_{0}$ in the
$64\times8$ MIMO system ($16\times4$ UPA) and the $16\times8$ MIMO
system ($16\times1$ ULA), respectively. The NMSE at the $i\text{th}$
turbo iteration is calculated by
\begin{equation}
\mathsf{NMSE}=\frac{1}{\Theta}\sum_{\theta=1}^{\Theta}\frac{1}{MN}\sum_{m=1}^{M}\sum_{n=1}^{N}\frac{\sum_{l=1}^{L}\bigl|h_{mnl}-\hat{h}_{mnl}^{(i)}\bigr|^{2}}{\sum_{l=1}^{L}\left|h_{mnl}\right|^{2}},
\end{equation}
where $\Theta$ is the number of Monte Carlo runs. It is shown that
the NMSE of the proposed EP-QA-L outperforms other algorithms including
the BP-GA, the BP-MF-M, and the BP-MF (which is evaluated only in
the $16\times8$ MIMO system due to complexity issue). It is also
shown that, compared with the EP-QA using oracle channel PDP, the
EP-QA-L is slightly degraded only in the low region of $E_{b}/N_{0}$.
The NMSE of BP-MF-M is higher than that of all other algorithms at
the point that the number of turbo iterations is 15.

Fig. \ref{fig:MSE_64_8_64QAM_Itera} and Fig. \ref{fig:MSE_16_8_16QAM_Itera}
present the NMSE performance with increasing number of turbo iterations.
In the high region of $E_{b}/N_{0}$, it can be seen that 10 turbo
iterations are enough for all the algorithms to achieve convergence.
In the low $E_{b}/N_{0}$ region, the EP-QA-L (and the EP-QA with
oracle PDP) can uniformly improve the NMSE performance by increasing
the number of turbo iterations, but other algorithms can't.

\begin{figure}
\centering\includegraphics[width=3.5in]{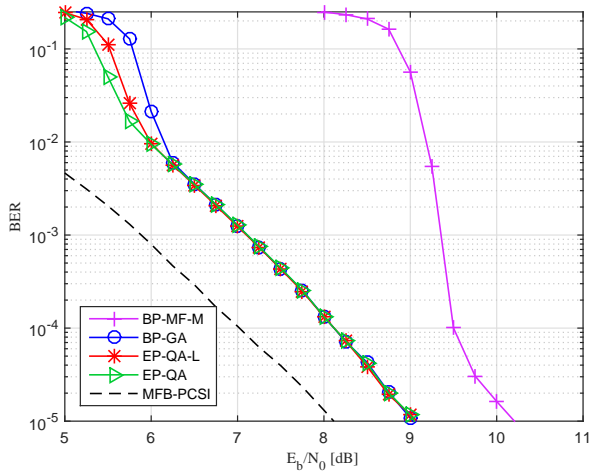}\caption{BER versus $E_{b}/N_{0}$ in the $64\times8$ MIMO system with 16QAM. }

\label{fig:BER_64_8_16QAM}
\end{figure}

\begin{figure}
\centering\includegraphics[width=3.5in]{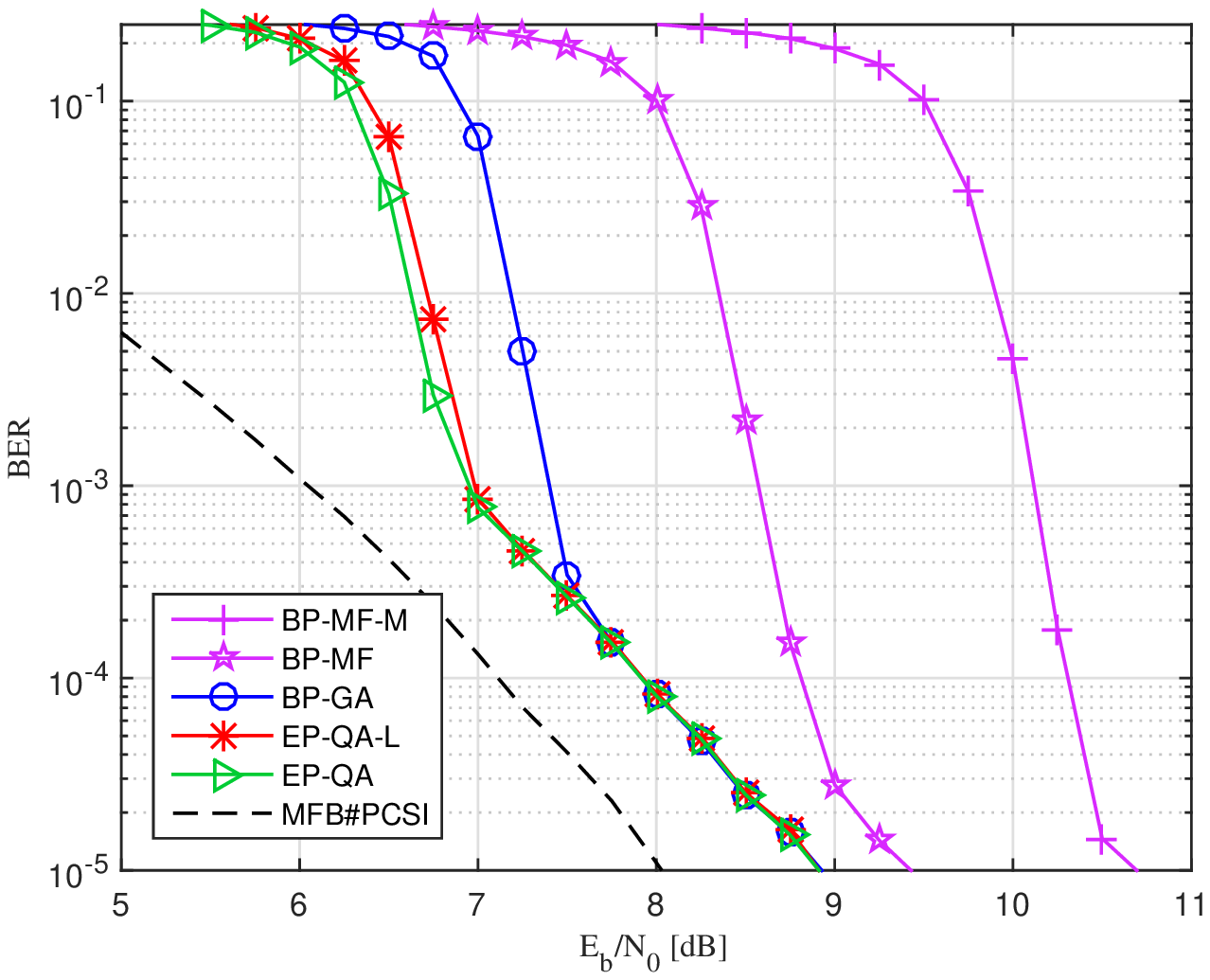}\caption{BER versus $E_{b}/N_{0}$ in the $16\times8$ MIMO system with 16QAM. }

\label{fig:BER_16_8_16QAM}
\end{figure}

\begin{figure}
\subfloat[BP-MF-M.]{\centering\includegraphics[width=3.25in]{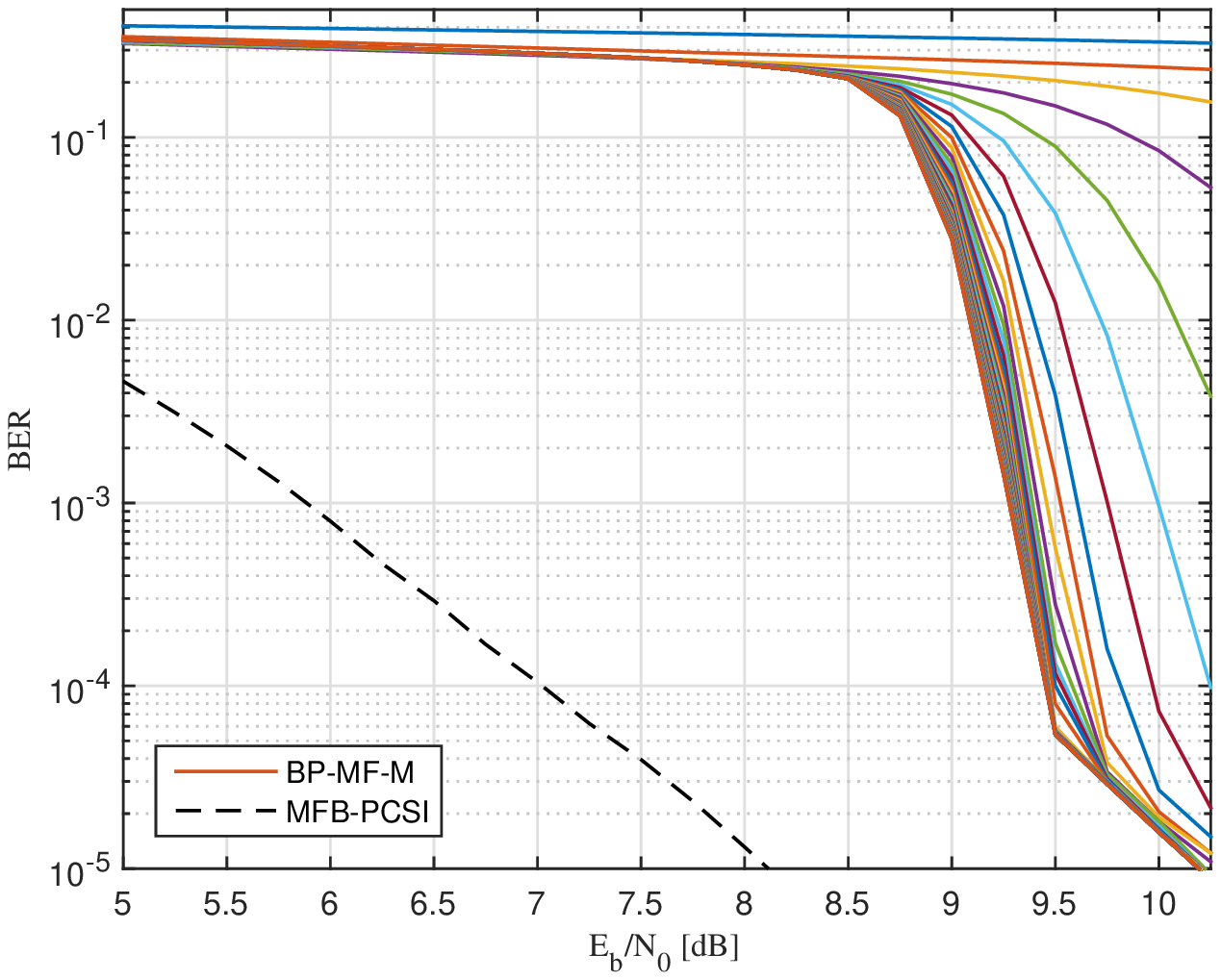}}\subfloat[BP-GA.]{\centering\includegraphics[width=3.25in]{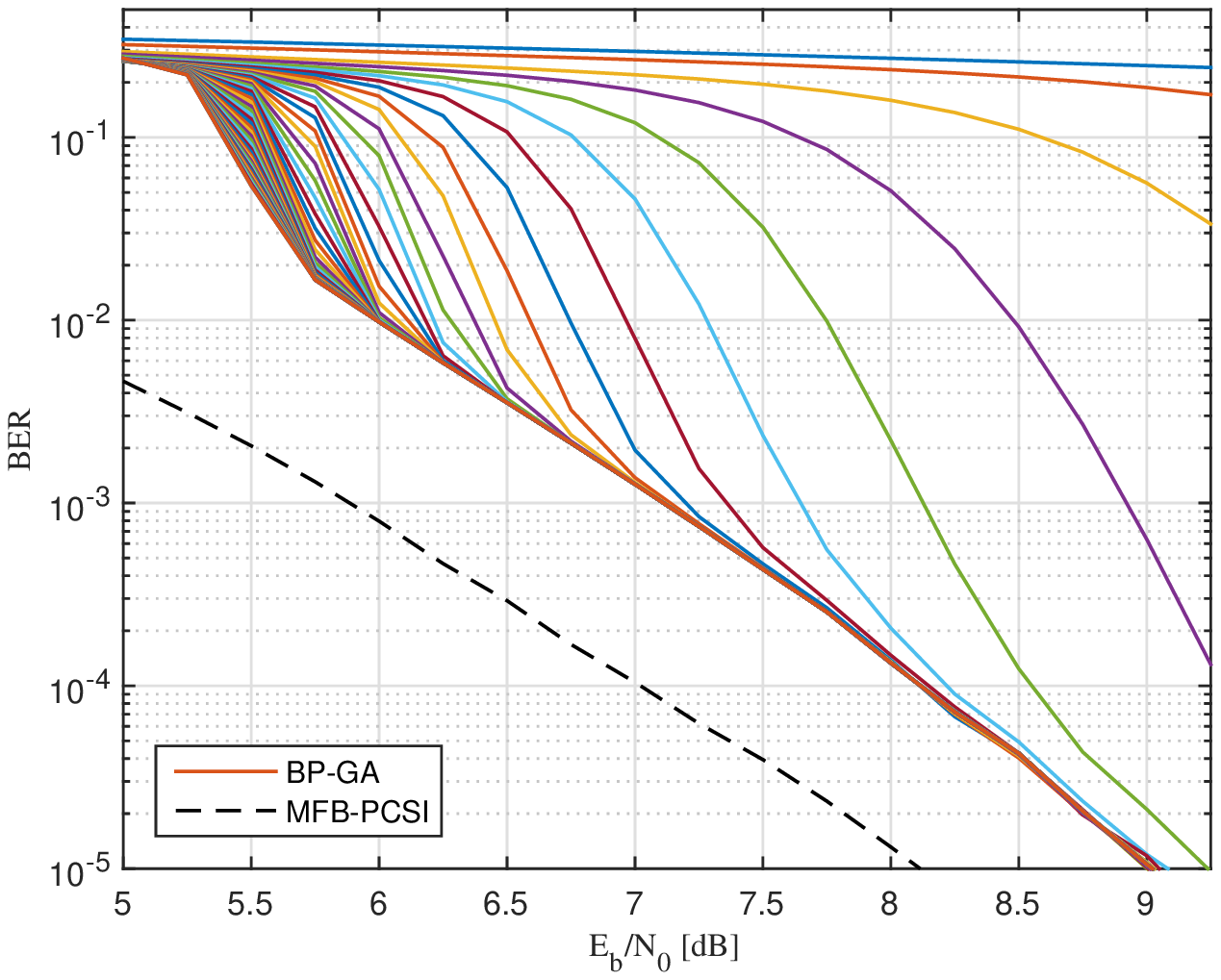}

}

\subfloat[EP-QA-L.]{\centering\includegraphics[width=3.25in]{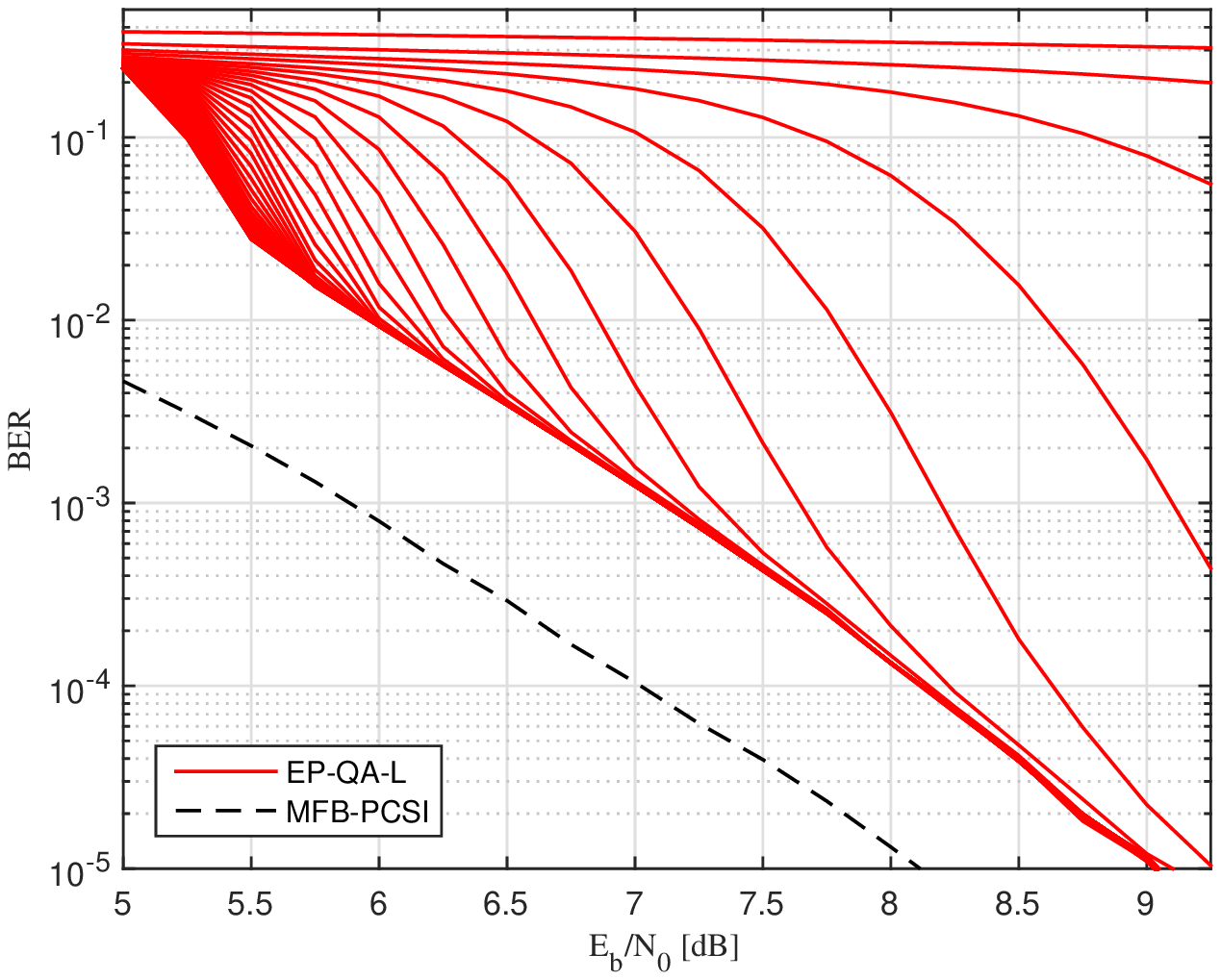}}\subfloat[EP-QA.]{\centering\includegraphics[width=3.25in]{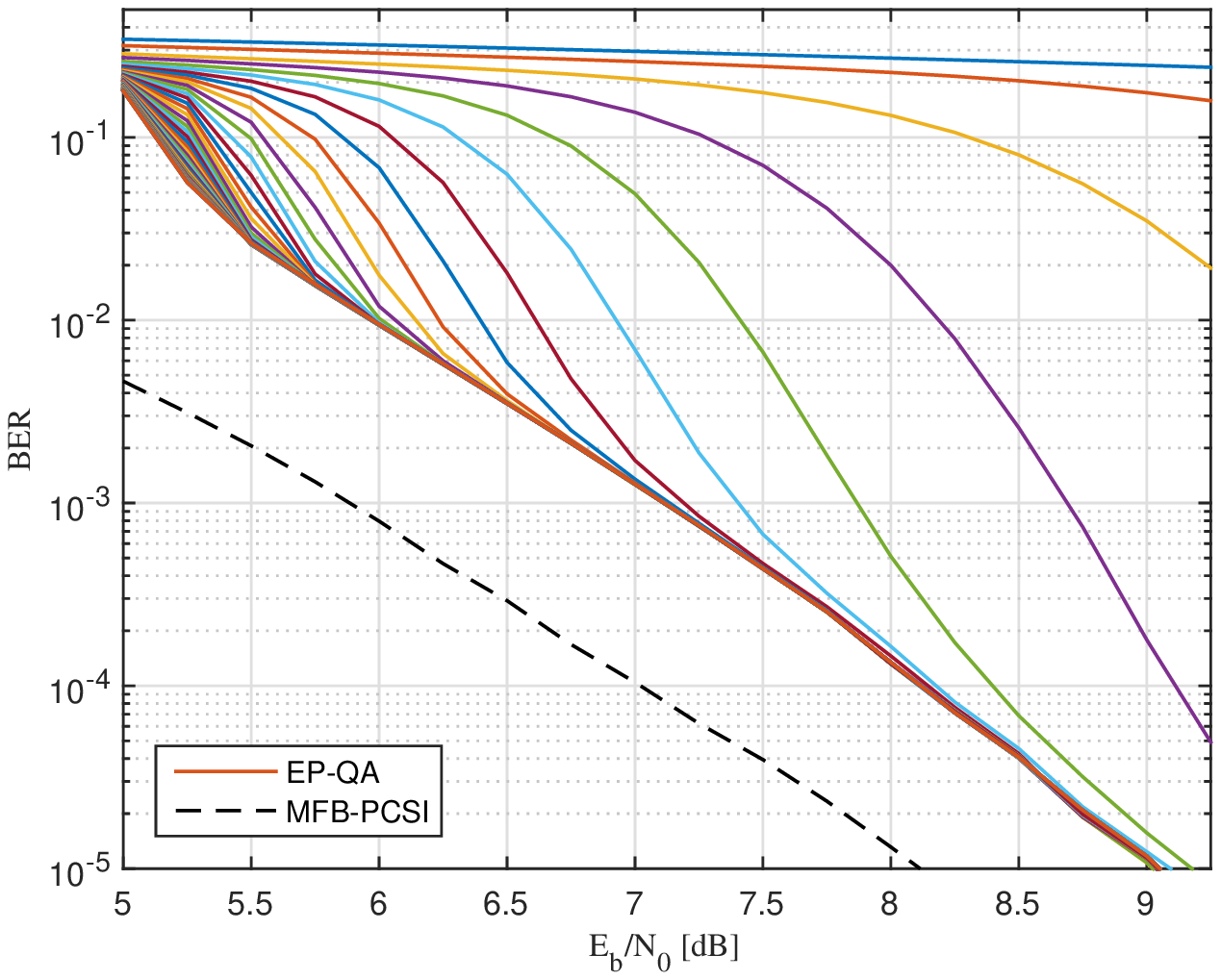}

}

\caption{BER versus $E_{b}/N_{0}$ in the $64\times8$ MIMO system with 16QAM. }
\label{fig:BER_64_8_16QAM_Itera}
\end{figure}

\begin{figure}
\subfloat[BP-MF-M.]{\centering\includegraphics[width=3.25in]{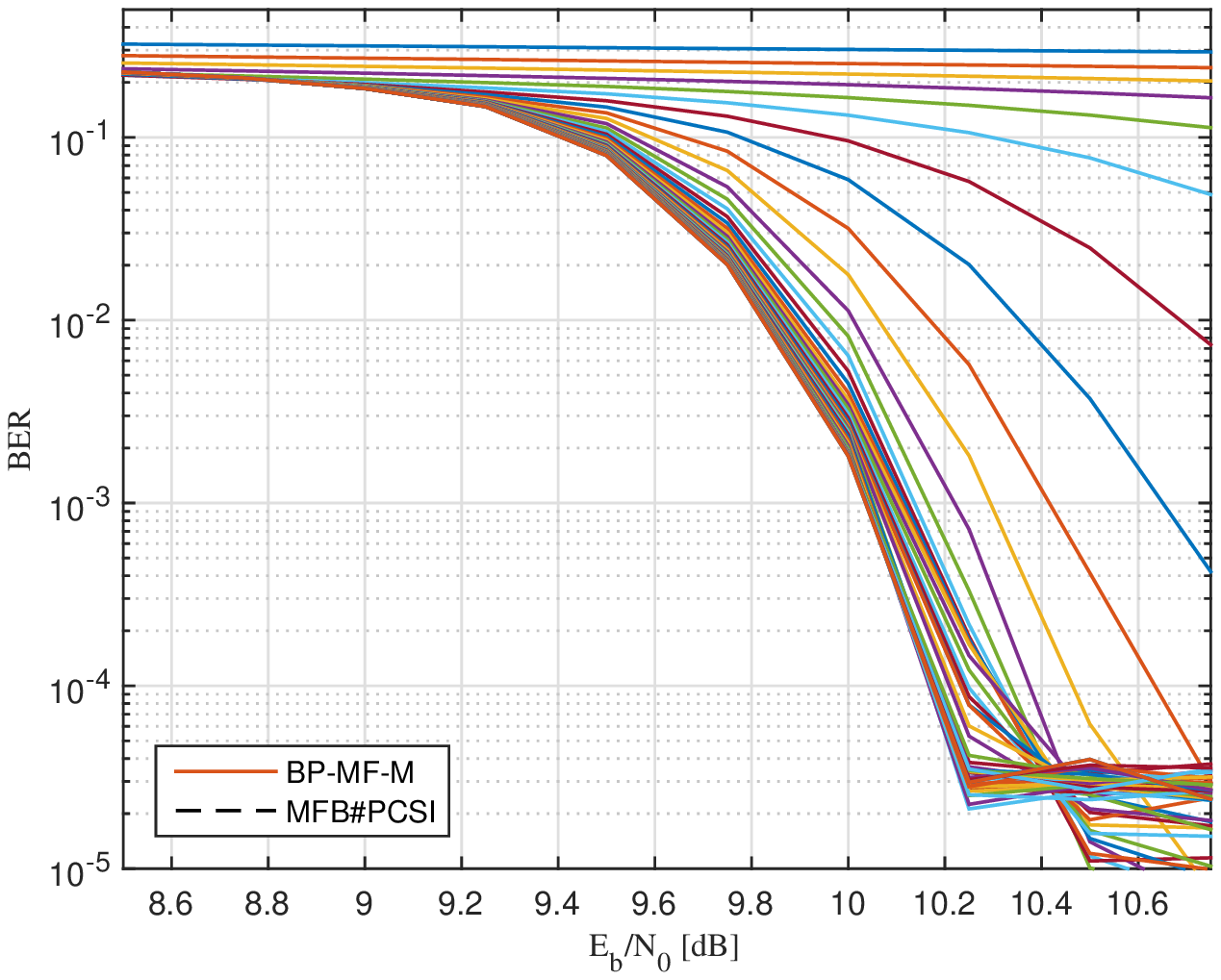}}\subfloat[BP-MF.]{\centering\includegraphics[width=3.25in]{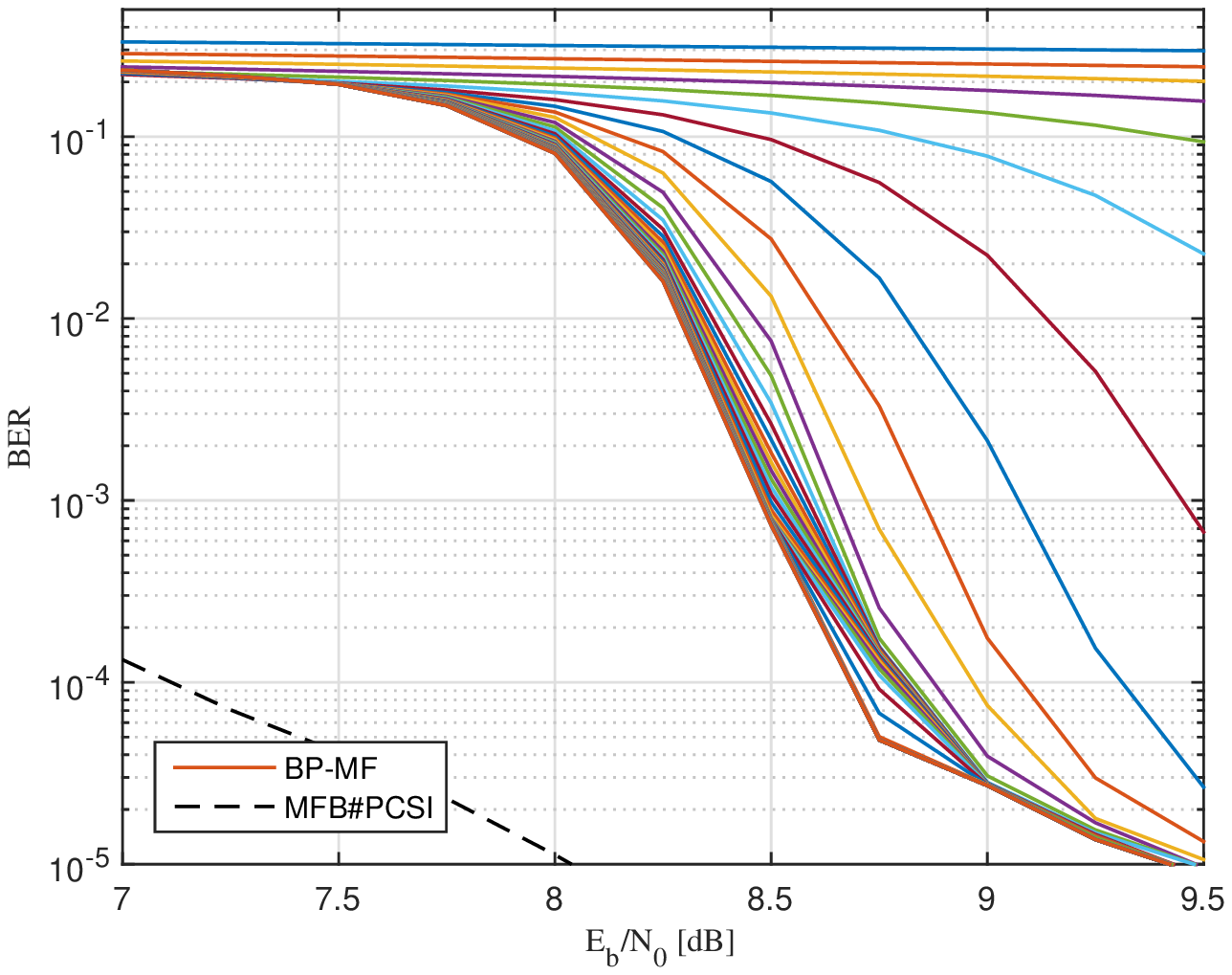}}

\subfloat[BP-GA.]{\centering\includegraphics[width=3.25in]{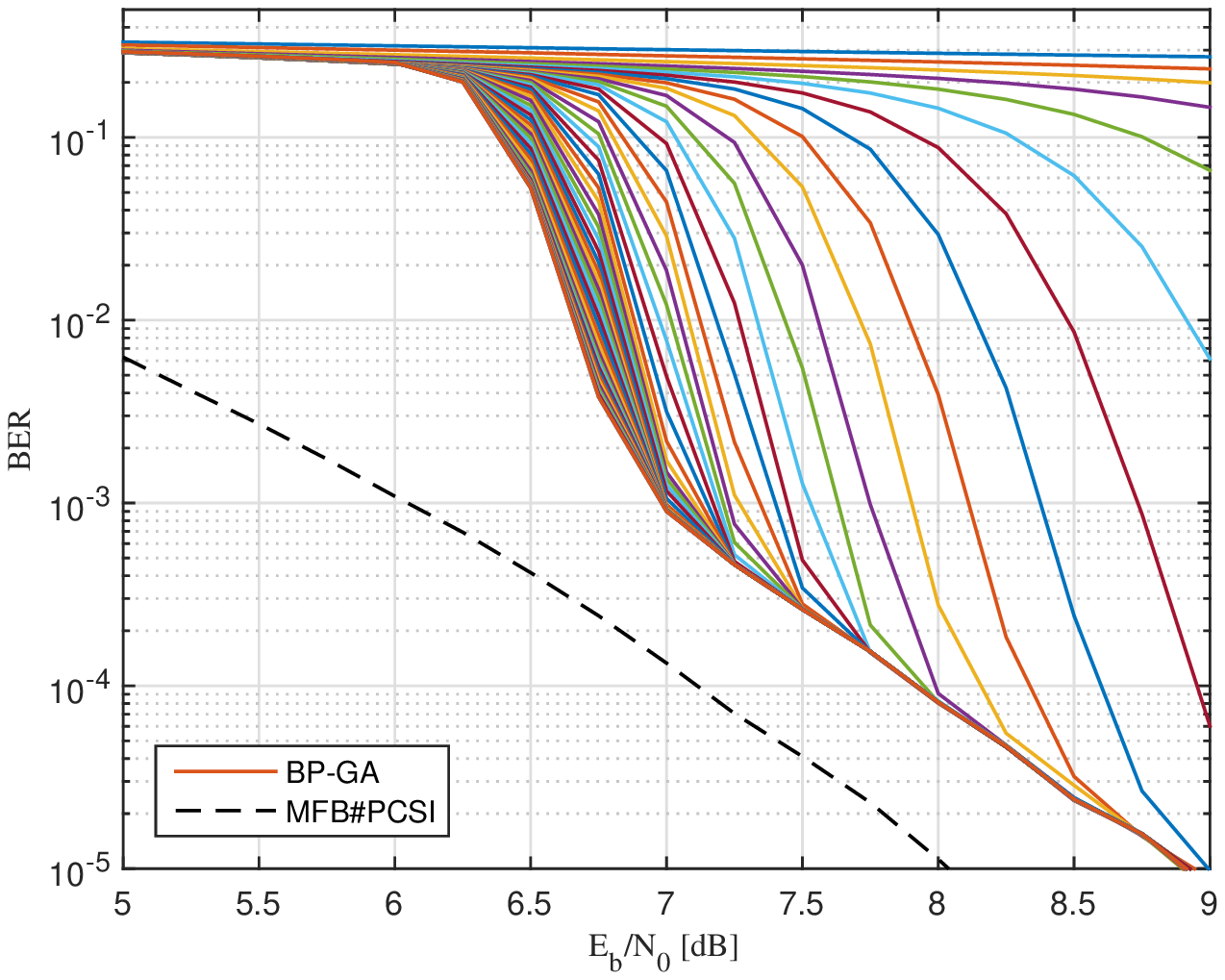}}\subfloat[EP-QA.]{\centering\includegraphics[width=3.25in]{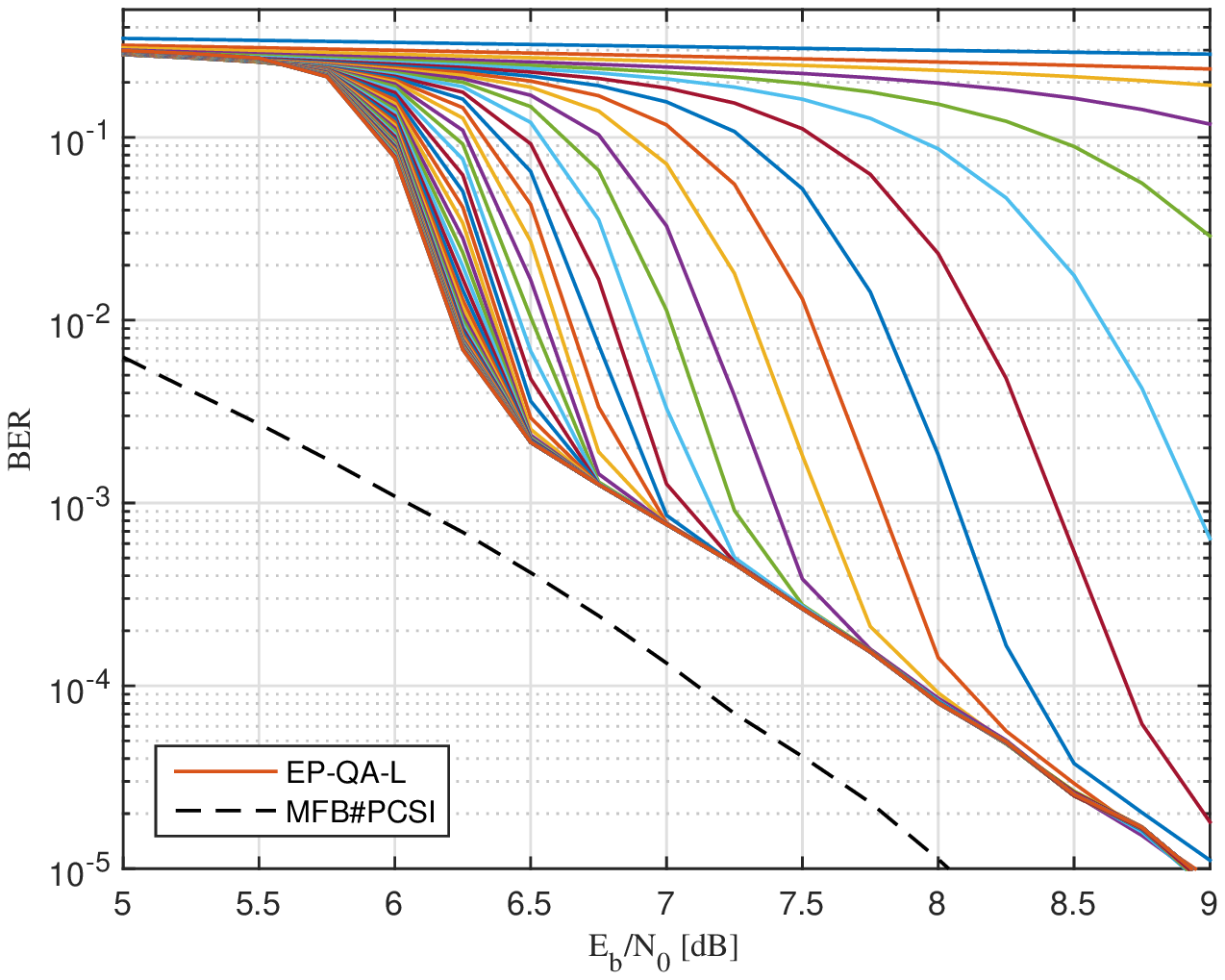}}

\centering\subfloat[EP-QA-L.]{\centering\includegraphics[width=3.25in]{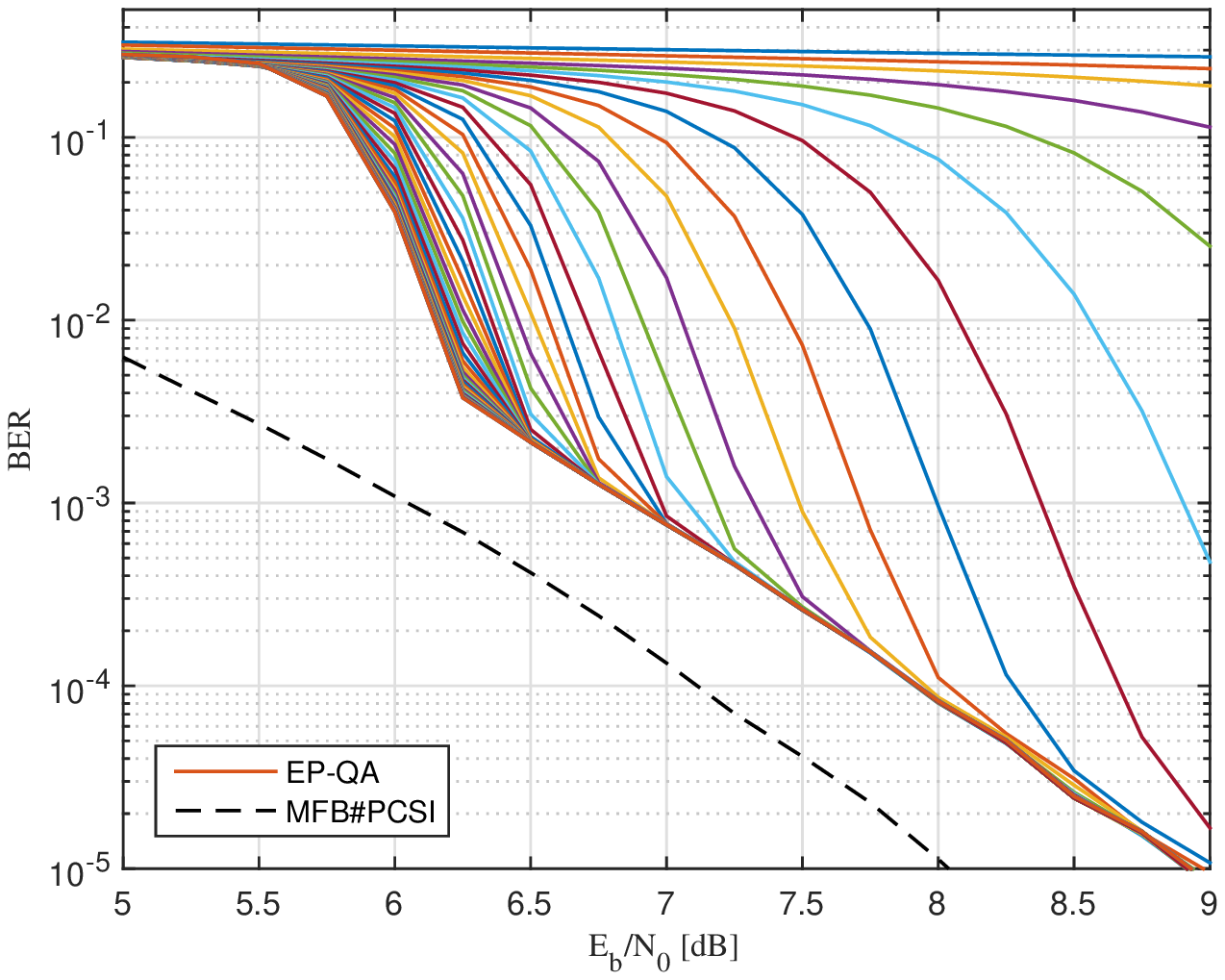}}

\caption{BER versus $E_{b}/N_{0}$ in the $16\times8$ MIMO system with 16QAM. }
\label{fig:BER_16_8_16QAM_Itera}
\end{figure}

\subsection{$BER$ Versus $E_{b}/N_{0}$}

Fig. \ref{fig:BER_64_8_16QAM} and Fig. \ref{fig:BER_16_8_16QAM}
show the BER performance versus $E_{b}/N_{0}$ in the $64\times8$
MIMO system and the $16\times8$ MIMO system, respectively. The BP-GA,
the EP-QA, and the EP-QA-L achieve the same performance that is about
$0.9\,\text{dB}$ away from the MFB-PCSI at $\text{BER}=10^{-5}$,
but the BP-MF-M is about $2.0\,\text{dB}$ away from the MFB-PCSI
in the $64\times8$ MIMO system and $2.6\,\text{dB}$ away from the
MFB-PCSI in the $16\times8$ MIMO system. In the $16\times8$ MIMO
system, even the BP-MF with much higher complexity is still inferior
to the EP-QA-L about $0.5\,\text{dB}$ at $\text{BER}=10^{-5}$.

Fig. \ref{fig:BER_64_8_16QAM_Itera} and Fig. \ref{fig:BER_16_8_16QAM_Itera}
show the BER performance versus $E_{b}/N_{0}$ with increasing number
of turbo iterations. In the $64\times8$ MIMO system, to converge
at $\text{BER}=10^{-5}$, the BP-GA, the EP-QA, and the EP-QA-L need
about 7 turbo iterations and the BP-MF-M needs about 12 turbo iterations.
In the $16\times8$ MIMO system, the BP-GA, the EP-QA, and the EP-QA-L
need 9 turbo iterations to converge at $\text{BER}=10^{-5}$ and the
BP-MF needs about 12 turbo iterations, while the performance of BP-MF-M
is somewhat unstable.

\section{Conclusion\label{sec:Conclusion}}

In this paper, we presented a message-passing receiver for joint channel-estimation
and decoding in the 3D massive MIMO systems transmitting over frequency-selective
block fading channels. Expectation propagation with quadratic approximation
was derived to deal with the decoupling of channel coefficients and
data symbols, and a low-complexity Gaussian message-passing algorithm
was applied for the channel estimation. It was verified through simulations
that in the 3D massive MIMO systems our proposed algorithm could approach
to the MFB with limited loss and its complexity is very low.

\section*{Acknowledgment}

The authors would like to gratefully acknowledge Prof. P. Schniter
for valuable suggestions, especially on the issue of negative variances
occurred in the EP algorithm.

\appendix{

Using the so-called\emph{ Wirtinger calculus }\cite{van1994complex,kreutz2009complex},
a real function of $z\in\mathbb{C}$, $\tau>0$ and $u\in\mathbb{C}$
is defined by
\begin{equation}
\mathcal{H}\left(\vec{\boldsymbol{z}},\tau,\vec{u}\right)=\frac{\left|z\right|^{2}}{\tau}+\mathsf{ln}\tau+\frac{\left|u\right|^{2}}{\nu},
\end{equation}
where the conjugate coordinates $\vec{\boldsymbol{z}}$ and $\vec{\boldsymbol{u}}$
are defined by $\vec{\boldsymbol{z}}\triangleq\left[z,z^{*}\right]^{\mathsf{T}}$
and $\vec{\boldsymbol{u}}\triangleq\left[u,u^{*}\right]^{\mathsf{T}}$
respectively, and $\nu>0$ is a constant. For the function $\mathcal{H}\left(\vec{\boldsymbol{z}},\tau,\vec{u}\right)$,
some of its partial derivations are given by
\begin{align}
\frac{\partial\mathcal{H}}{\partial\vec{\boldsymbol{z}}} & \triangleq\left[\frac{\partial\mathcal{H}}{\partial z},\frac{\partial\mathcal{H}}{\partial z^{*}}\right]=\left[\frac{z^{*}}{\tau},\frac{z}{\tau}\right],\\
\frac{\partial^{2}\mathcal{H}}{\partial\vec{\boldsymbol{z}}\partial\vec{\boldsymbol{z}}} & \triangleq\left[\begin{array}{cc}
\frac{\partial}{\partial z}\left(\frac{\partial\mathcal{H}}{\partial z}\right)^{*} & \frac{\partial}{\partial z^{*}}\left(\frac{\partial\mathcal{H}}{\partial z}\right)^{*}\\
\frac{\partial}{\partial z}\left(\frac{\partial\mathcal{H}}{\partial z^{*}}\right)^{*} & \frac{\partial}{\partial z^{*}}\left(\frac{\partial\mathcal{H}}{\partial z^{*}}\right)^{*}
\end{array}\right]=\left[\begin{array}{cc}
\frac{1}{\tau} & 0\\
0 & \frac{1}{\tau}
\end{array}\right],\\
\frac{\partial\mathcal{H}}{\partial\tau} & =\frac{1}{\tau}-\frac{\left|z\right|^{2}}{\tau^{2}},\\
\frac{\partial\mathcal{H}}{\partial\vec{\boldsymbol{u}}} & \triangleq\left[\frac{\partial\mathcal{H}}{\partial u},\frac{\partial\mathcal{H}}{\partial u^{*}}\right]=\left[\frac{u*}{\nu},\frac{u}{\nu}\right],\\
\frac{\partial^{2}\mathcal{H}}{\partial\vec{\boldsymbol{u}}\partial\vec{\boldsymbol{u}}} & \triangleq\left[\begin{array}{cc}
\frac{\partial}{\partial u}\left(\frac{\partial\mathcal{H}}{\partial u}\right)^{*} & \frac{\partial}{\partial u*}\left(\frac{\partial\mathcal{H}}{\partial u}\right)^{*}\\
\frac{\partial}{\partial u}\left(\frac{\partial\mathcal{H}}{\partial u^{*}}\right)^{*} & \frac{\partial}{\partial u^{*}}\left(\frac{\partial\mathcal{H}}{\partial u^{*}}\right)^{*}
\end{array}\right]=\left[\begin{array}{cc}
\frac{1}{\nu} & 0\\
0 & \frac{1}{\nu}
\end{array}\right].
\end{align}
Up to the second order, the power series expansion of $\mathcal{H}\left(\vec{\boldsymbol{z}},\tau,\vec{\boldsymbol{u}}\right)$
at the point $\left(\vec{\boldsymbol{z}}_{0},\tau_{0},\vec{\boldsymbol{u}}_{0}\right)$
is given by \cite{kreutz2009complex}
\begin{align}
\mathcal{H}\left(\vec{\boldsymbol{z}},\tau,\vec{\boldsymbol{u}}\right) & \approx\mathcal{H}\left(\vec{\boldsymbol{z}}_{0},\tau_{0},\vec{\boldsymbol{u}}_{0}\right)+\frac{\partial\mathcal{H}}{\partial\vec{\boldsymbol{z}}_{0}}\Delta\vec{\boldsymbol{z}}+\frac{\partial\mathcal{H}}{\partial\tau_{0}}\Delta\tau+\frac{\partial\mathcal{H}}{\partial\vec{\boldsymbol{u}}_{0}}\Delta\vec{\boldsymbol{u}}\nonumber \\
 & \hspace{1em}+\frac{1}{2}\left(\Delta\vec{\boldsymbol{z}}\right)^{\mathsf{H}}\frac{\partial^{2}\mathcal{H}}{\partial\vec{\boldsymbol{z}}_{0}\partial\vec{\boldsymbol{z}}_{0}}\Delta\vec{\boldsymbol{z}}+\frac{1}{2}\left(\Delta\vec{\boldsymbol{u}}\right)^{\mathsf{H}}\frac{\partial^{2}\mathcal{H}}{\partial\vec{\boldsymbol{u}}_{0}\partial\vec{\boldsymbol{u}}_{0}}\Delta\vec{\boldsymbol{u}}\nonumber \\
 & =\mathcal{H}\left(\vec{\boldsymbol{z}}_{0},\tau_{0},\vec{\boldsymbol{u}}_{0}\right)+2\Re\left\{ \frac{z_{0}^{*}}{\tau_{0}}\Delta z+\frac{u_{0}^{*}}{_{\nu}}\Delta u\right\} \nonumber \\
 & \hspace{1em}-\frac{\left|z_{0}\right|^{2}}{\tau_{0}^{2}}\Delta\tau+\frac{1}{\tau_{0}}\Delta\tau+\frac{1}{\tau_{0}}\left|\Delta z\right|^{2}+\frac{1}{\nu}\left|\Delta u\right|^{2},\label{eq:=00007BH=00007D(=00007B=00007Bu=00007D=00007D,=00007B=00007Bv=00007D=00007D)}
\end{align}
where $\Delta\vec{\boldsymbol{z}}\triangleq\vec{\boldsymbol{z}}-\vec{\boldsymbol{z}}_{0}$,
$\Delta\tau\triangleq\tau-\tau_{0}$ and $\Delta\vec{\boldsymbol{u}}\triangleq\vec{\boldsymbol{u}}-\vec{\boldsymbol{u}}_{0}$
}\bibliographystyle{IEEEtran}
\bibliography{IEEEabrv,Mybib}

\end{document}